\documentclass[a4paper,11pt]{article}
\pdfoutput=1

\usepackage{jheppub1}
\usepackage[T1]{fontenc}
\usepackage{amssymb}
\usepackage{graphicx}
\usepackage{amsmath}
\usepackage{tensor}
\usepackage{hyperref}
\usepackage{epstopdf}
\usepackage{extarrows}
\usepackage{xcolor}
\usepackage{tensor}
\usepackage{subfigure}

\newcommand{\td}{\text{d}}

\def\x {\hat{x}}
\def\p {\hat{p}}
\def\O {\hat{O}}
\def\U {\hat{U}}
\def\C {\mathcal{C}}

\def\Fi {\text{Fi}}

\def\pD {\mathcal{D}}
\def\Tr {\text{Tr}}

\begin{document}
\title{To be unitary-invariant or not?: a simple but non-trivial proposal for the complexity between states in quantum mechanics/field theory}
\author[a]{Run-Qiu Yang,}
\author[b,c]{Yu-Sen An,}
\author[d]{Chao Niu,}
\author[e]{Cheng-Yong Zhang,}
\author[f]{Keun-Young Kim}

\emailAdd{aqiu@kias.re.kr}
\emailAdd{anyusen@itp.ac.cn}
\emailAdd{chaoniu09@gmail.com}
\emailAdd{zhangchengyong@fudan.edu.cn}
\emailAdd{fortoe@gist.ac.kr}

\affiliation[a]{Quantum Universe Center, Korea Institute for Advanced Study, Seoul 130-722, Korea}
\affiliation[b]{Institute of Theoretical Physics, Chinese Academy of Science, Beijing 100190, China}
\affiliation[c]{School of physical Science, University of Chinese Academy of Science, Beijing 100049, China}
\affiliation[d]{Department of Physics and Siyuan Laboratory, Jinan University, Guangzhou 510632, China}
\affiliation[e]{Department of Physics and Center for Field Theory and Particle Physics, Fudan University, Shanghai 200433, China}
\affiliation[f]{ School of Physics and Chemistry, Gwangju Institute of Science and Technology,
Gwangju 61005, Korea
}

\abstract{
We make comments on some  shortcomings of the non-unitary-invariant and non-bi-invariant complexity in {\it quantum mechanics/field theory} and argue that the unitary-invariant and bi-invariant complexity is still a competitive candidate in {\it quantum mechanics/field theory},  contrary to {\it quantum circuits in quantum computation}.
Based on the unitary-invariance of the complexity and intuitions from the holographic complexity, we propose a novel complexity formula between two states. Our proposal shows that i) the complexity between certain states in two dimensional CFTs is given by the Liouville action, which is compatible with the path-integral complexity; ii) it also gives natural interpretation for both the CV and CA holographic conjectures and identify what the reference states are in both cases. Our proposal explicitly produces the conjectured time dependence of the complexity: linear growth in chaotic systems. Last but not least, we present  interesting relations between the complexity and the Lyapunov exponent: the Lyapunov exponent is proportional to the complexity growth rate in linear growth region.
}

\maketitle

\noindent

\section{Introduction}
Recently, the concepts in quantum information theory have been applied to investigate the theory of gravity and black holes. In particular, a concept named ``complexity'', which comes from the quantum circuit complexity in quantum information theory,  was introduced for the study of the black hole  interior. The complexity in quantum circuits can be defined for both operators and states. Roughly speaking, the complexity of an operator is the minimal number of required gates\footnote{The gates are basic building blocks to construct the quantum circuit.} when we use quantum circuits to simulate it; the complexity between a reference state and a target state is the minimal number of required gates when we use quantum circuits to transform a  reference state to  a target state.

The motivation to introduce the complexity into the black holes physics was to understand about the fire-wall model of the black hole~\cite{Harlow:2013tf} and the growth rate of the Einstein-Rosen bridge for the AdS black holes~\cite{Stanford:2014jda,Susskind:2014rva,Susskind:2014rva2}.  Refs.~\cite{Stanford:2014jda} and \cite{Brown:2015bva} proposed two holographic conjectures to compute the complexity for some particular quantum states which are dual to boundary time slices of an eternal asymptotic AdS black hole. They are called the complexity-volume (CV) conjecture and the complexity-action (CA) conjecture.

The CV conjecture states that the complexity is proportional to the maximum volume of time-like hypersurfaces. Suppose $t_L$ and $t_R$ are  two time slices at the left and right boundaries of an external asymptotic AdS black hole. Then the CV conjecture is given by
\begin{equation}\label{CVeq1}
  \C=\max_{\partial\Sigma=t_L\cup t_R}\frac{\text{Vol}(\Sigma)}{G_N\ell}\,,
\end{equation}
where $\Sigma$ is a spacelike surface which  connects the time slices $t_L$ and $t_R$ of two boundaries, $G_N$ is the Newton's gravity constant and $\ell$ is a length scale.
The CA conjecture states that the complexity associated to two boundary time slices is given by the on-shell action in the Wheeler-DeWitt (WdW) patch
\begin{equation}\label{CAeq1}
  \C=\frac{S_{\text{WdW,on-shell}}}{\pi\hbar}\,.
\end{equation}
The WdW patch is the closure of all spacelike surfaces which connect $t_L$ and $t_R$.

Many works have been done to study the properties of the conjectures~\eqref{CVeq1} and \eqref{CAeq1}: the time-evolution of the holographic complexity in the CV or CA conjectures~\cite{Carmi:2017jqz,Kim:2017qrq,An:2018dbz}, the action growth rate and the Lloyd's bound in various gravity systems~\cite{Cai:2016xho,Yang:2016awy,Pan:2016ecg,Alishahiha:2017hwg,An:2018xhv,Jiang:2018pfk,Jiang:2018sqj, Yang:2019gce, HosseiniMansoori:2018gdu}, the UV divergent structures of the holographic complexity~\cite{Chapman:2016hwi,Kim:2017lrw}, the quench effects in the holographic complexity~\cite{Moosa:2017yvt,Chen:2018mcc,Fan:2018xwf} and so on. Besides these two conjectures, other conjectures for the complexity were also proposed in holography for different systems and purposes (see, for example, Refs~\cite{Alishahiha:2015rta,Ben-Ami:2016qex,Couch:2016exn,Caputa:2017urj,Caputa:2017yrh,Fan:2018wnv,Fan:2019mbp}).

Though all these results give us some understandings about the holographic complexity, a few of fundamental questions are still unsolved. The most important one is what the reference states in the CV and CA conjectures are. Both the CV and CA conjectures are expected to describe the complexity between states, which will be meaningful only if both the reference state and the target state are identified clearly. The target state is dual to the thermofield double (TFD) state associated with time slices at the boundary~\cite{Maldacena:2001kr}. However, the reference state is unclear in the statements of both the CV and CA conjectures.

The other question is how to understand different behaviors of the time-evolution in the CV and CA conjectures. Though both the CV and CA conjectures shows that the complexity grows linearly at late time limit, they show different behaviors at early time. In the CV conjecture, the complexity  grows as $t^2$ at early time while, in the CA conjecture, it first keeps constant and then suddenly obtain a negative infinite growth rate after a certain time. See Refs.~\cite{Carmi:2017jqz,Kim:2017qrq} for more detailed discussions about the time evolutions of the complexity in the CV and CA conjectures. This difference may imply that two conjectures describe two different complexities in field theory rather than the previous expectation that they both describe the complexity between the TFD state and an unkonwn ``simple'' reference state.

Compared with much progress on the complexity in gravity side, the exact meaning and a well-proposed definition of the complexity in quantum field theory is still incomplete.\footnote{Recently, there have been many attempts to generalize the concept of complexity of discrete quantum circuit to continuous systems such as ``complexity geometry''~\cite{Susskind:2014jwa,Brown:2016wib,Brown:2017jil} based on ~\cite{Nielsen1133,Nielsen:2006:GAQ:2011686.2011688,Dowling:2008:GQC:2016985.2016986}, Fubini-study metric~\cite{Chapman:2017rqy}, and path-integral optimization~\cite{Caputa:2017urj,Caputa:2017yrh,Bhattacharyya:2018wym,Takayanagi:2018pml}. See also \cite{Hashimoto:2017fga,Hashimoto:2018bmb,Flory:2018akz,Flory:2019kah,Belin:2018fxe,Belin:2018bpg}. In particular, the complexity geometry is
the most studied. See for exampe~\cite{Jefferson:2017sdb,Yang:2017nfn,Reynolds:2017jfs,Kim:2017qrq,Khan:2018rzm,Hackl:2018ptj,Yang:2018nda,Yang:2018tpo,Alves:2018qfv,Magan:2018nmu,Caputa:2018kdj,Camargo:2018eof,Guo:2018kzl,Bhattacharyya:2018bbv, Jiang:2018gft,Camargo:2018eof,Chapman:2018hou,Ali:2018fcz,Chapman:2018dem}. }  In quantum circuits, the complexity is defined in the  discrete and finite Hilbert spaces. The definition of the complexity in terms of quantum gates may be ideal for computer science, but not for field theory, a continuous system.

The first attempt to find a generalization of the circuit complexity to continuous systems was proposed by Nielsen et al.~\cite{Nielsen1133,Nielsen:2006:GAQ:2011686.2011688,Dowling:2008:GQC:2016985.2016986}. They constructed a continuum approximation of the circuit complexity which involves the geodesic distance in a certain geometry called  ``complexity geometry''. The recent works such as Refs.~\cite{Brown:2017jil,Jefferson:2017sdb,Yang:2017nfn,Chapman:2017rqy,Khan:2018rzm,Camargo:2018eof,Chapman:2018hou} followed Nielsen's right-invariant complexity geometry to define the complexity between states.

However, all these works {reach a conclusion or assumption:} the complexity is \emph{not invariant} if we make the same unitary transformation for both the reference and target state.  We will call this property ``non-unitary invariant.'' This implies that the complexity is \textit{bases-dependent} but there is no clear physical principle  to choose a unique ``favored''  base. In order to obtain the desired results, the bases and corresponding metric components need to be chosen carefully by hand rather than determined by physical principles.  Thus, it is hard to say whether the results in these approaches describe the properties of the physical systems or the properties of such artificial choices.


Different from the geometrization method of Nielsen's, Refs.~\cite{Caputa:2017urj,Caputa:2017yrh,Bhattacharyya:2018wym} proposed the ``path-integral complexity'' to describe the complexity between  the field operator eigenstate and the ground state of a 2-dimensional conformal field theory (CFT). It states that the complexity can be given by the on-shell Liouville action. This is based on the tensor network renormalizations~\cite{PhysRevLett.115.180405} in constructing the ground state. Ref.~\cite{Czech:2017ryf} also proved the Einstein's equation in 2+1 dimensional case could be obtained by minimizing such a complexity. Recently, Ref.~\cite{Camargo:2019isp} offers a viewpoint to connect the path integral complexity and circuit complexity and tries to fill up the gap between these two different proposals in field theory. The  ``path-integral complexity'' has an essential difference compared with the geometrization method of Nielsen's: it is\textit{ bases-independent} and so is \textit{unitary invariant}.


The first goal of our paper is to show that the complexity should be unitary-invariant, contrary to the complexity in the quantum circuits in quantum computation science, which we will call `real quantum circuits'.
We will first review the main features of the non-unitary-invariant complexity and explain four crucial shortcomings from the viewpoint of quantum mechanics/field theory and holographic conjectures. We emphasize that these shortcomings do not arise in real quantum circuits so we do not claim that the complexity is in general unitary-invariant; It is better to be unitary invariant in quantum mechanics/field theory.

The second goal is to propose a novel {\it unitary invariant} complexity formula between two states $|\psi_1\rangle$ and $|\psi_2\rangle$,
\begin{equation}\label{defholoCeq200}
  \C(|\psi_1\rangle,|\psi_2\rangle)=- \ln|\langle\psi_1|\psi_2\rangle|^2\,.
\end{equation}
We will show this simple formula implies many interesting consequences.
Firstly, It proves that the states complexity in 2D CFTs is given by the Liouville action, which is consistent with the path-integral complexity.  Secondly, it gives natural explanations for both the CV and CA conjectures.  In  particular, it clarifies what the target and reference states are in the CV and CA conjectures. In other words, our proposal answers two aforementioned unsolved questions.

The third goal is to show that our proposal can demonstrate the widely accepted but not-yet-proven time dependence of the complexity: nearly linear growth before the saturation in chaotic systems.
Last but not least, we present very interesting relations between the complexity and the Lyapunov exponent,
\begin{equation}\label{complexL110022}
  \C(t)={\lambda}_Lt+\cdots\,,
\end{equation}
and the saturation time (the time at the end of the linear growth) and the Lyapunov exponent,
\begin{equation}\label{logtime110022}
t_{cl}:=-\frac{1}{2{\lambda}_L}\ln(c_1\hbar) \,,
\end{equation}
where $c_1$ is a model-dependent constant.
In Ref.~\cite{Yang:2019iav} we provide concrete simple examples supporting these theoretical predictions. Note that
it is often claimed~\cite{Brown:2017jil, Balasubramanian:2019wgd} that the complexity must be non-unitary-invariant because a unitary-invariant complexity cannot show \eqref{complexL110022} in an exponential time scale (for a chaotic system with $N$ degrees of freedom, it means $t\sim e^{N}$). This paper together with Ref.~\cite{Yang:2019iav}
give counter examples of this claim and support the possibility that the complexity may be unitary-invariant.


The paper is organized as follows: In section~\ref{reivewC}, we briefly review on non-unitary-invariant complexity. In section~\ref{prob-non-un}, we describes problems of non-unitary-invariant complexity. In section~\ref{commlocal}, we make some comments on the concept of ``locality'' {and its relation to complexity. In particular we clarify why we need to distinguish ``apparent locality'' and ``intrinsic locality''.} In section~\ref{C-states}, a novel unitary-invariant complexity formula is proposed and its implications are discussed. In section~\ref{app-chaos}, we apply our formula to chaotic systems and show it produces an expected time-dependent complexity. In section~\ref{answers1}, we made two comments on our proposal: difference from the Fubini-Study distance and applications to the TFD state. We conclude in section~\ref{summ}.

\section{Review on non-unitary-invariant complexity}\label{reivewC}
In this section, we first review the main motivations and features of the non-unitary-invariant complexity proposed by a few literatures, such as Refs.~\cite{Nielsen1133,Nielsen:2006:GAQ:2011686.2011688,Dowling:2008:GQC:2016985.2016986,Susskind:2014jwa,Brown:2017jil,Jefferson:2017sdb}.

Let us begin with the complexity in quantum circuits. In the language of quantum circuits, the fundamental observables are ``gates'' $g_i$, which are basic quantum circuits operating on a small number of qubits and are the building blocks of quantum circuits. By suitably arranging and connecting these gates, we can form a bigger quantum circuits which can be used to simulate a unitary operator $\U$. For example,\footnote{In general, the gates can be connected by more complicatedly ``\textit{graph}''. For simplicity, we only consider that all the gates are aligned in one line.}
\begin{equation}\label{timeoder1}
  \U=g_ng_{n-1}\cdots g_2g_1\,.
\end{equation}
Mathematically, it has been proven that, in a given truncation error, a set of finitely many different gates, which are called ``universal fundamental gates set'', can be used to simulate all unitary operators.  A universal fundamental gates set is not unique and infinitely many different universal fundamental gates sets can be used.  The complexity in quantum circuits is defined by the following two steps.

For a given a universal fundamental gates set $\mathcal{S}=\{g_1,g_2,\cdots,g_n\}$, we first define the complexity of a unitary operator $\U$ such that
\begin{equation}\label{defCforU1}
  \C(\U)=\min N\,,\quad \text{such that}\quad g_{i_N}g_{i_{N-1}}\cdots g_{i_2}g_{i_1}\approx\U\,,
\end{equation}
where $g_{i_N},g_{i_{N-1}}\cdots, g_{i_2},g_{i_1}\in\mathcal{S}$.
Roughly speaking, the complexity of $\U$ is the minimal required number of gates when we simulate $\U$ by gates in universal fundamental gates set $\mathcal{S}$. In quantum circuits, as all the gates are invertible, we have
\begin{equation}
g_i\in\mathcal{S}\iff g_i^{-1}\in\mathcal{S}\,,
\end{equation}
which means
\begin{equation}
\C(\U)=\C(\U^{-1})\,.
\end{equation}
The complexity between two quantum states then is defined by the minimal complexity of the operators which can transform one to the other,
\begin{equation}\label{cforstates1}
  \C(|\psi_1\rangle,|\psi_2\rangle)=\min\{\C(\U)|~\forall \U, s.t., |\psi_1\rangle=\U|\psi_2\rangle\}\,.
\end{equation}

{Note that this definition of the complexity in quantum circuits depends on the choice of a universal fundamental gates set. Thus, such a dependence may not reflect the intrinsic properties of operators and quantum states.  However, in the interest of the quantum computation based on quantum circuits, universal fundamental gates sets are part of the game, together with the quantum states. Furthermore, based on current technology, only a few of universal fundamental gates sets can be created by human, so the  dependence on the fundamental gates set will not be a big issue. However, if we want to study the complexity between states transformed by physical law not by the man-made circuits, this may be an issue.}

Nielsen and his collaborators first tried to generalize the complexity to  continuous systems~\cite{Nielsen1133,Nielsen:2006:GAQ:2011686.2011688} in a finite dimensional Hilbert space. In continuous systems, the universal fundamental gates sets are replaced by a Lie algebra $\mathfrak{g}$.  To construct a unitary operator $\U$, we need to choose a path-dependent generator $iH(s)\in\mathfrak{g}$ such that
\begin{equation}\label{condgsU}
  \U=\mathcal{P}\exp\int_0^1iH(s)\td s\,,
\end{equation}
where $\mathcal{P}$ means the path-order. Many literatures assume that this time-order is just the left product order, $\mathcal{P}=\overleftarrow{\mathcal{P}}$. However, we want to emphasize that  the {\it right} product order $\mathcal{P}=\overrightarrow{\mathcal{P}}$ is equally acceptable.

To define the complexity of the operator $\U$ in this set-up, we first introduce an inner product $\tilde{g}(\cdot,\cdot)$ for the Lie algebra $\mathfrak{g}$. With a bases $\{ie_I\}$ for the Lie algebra $\mathfrak{g}$, the inner product can be given by a Riemannian metric $\tilde{g}_{IJ}$ such that\footnote{{Though Nielsen in his original work~\cite{Nielsen:2006:GAQ:2011686.2011688} put the theory in a general Finsler geometry and also mentioned a kind of non-Riemannian metric, which is defined by a ``$F_1$'' norm (see Ref.~\cite{Nielsen:2006:GAQ:2011686.2011688} for details about $F_1$ norm), was most promising metric, many literatures, including Nielsen's work such as Ref.~\cite{Nielsen1133}, still used the Riemannian metric to defined complexity for simplicity.}}
\begin{equation}\label{metricgs1}
  \tilde{g}(H,H)=\tilde{g}_{IJ}Y^IY^J,~~~H=Y^Ie_I\,.
\end{equation}
If we choose  the left-order $\mathcal{P}=\overleftarrow{\mathcal{P}}$, an  arbitrary curve is given by
\begin{equation}\label{defcs1}
  c(s)=\overleftarrow{\mathcal{P}}\exp\int_0^siH(t)\td t\,.
\end{equation}
By defining the length $L[c]$ of the curve $c(s)$ as
\begin{equation}\label{curLs1}
  L[c]=\int_0^1\sqrt{\tilde{g}(H(s),H(s))}\td s=\int_0^1\sqrt{\tilde{g}_{IJ}Y^I(s)Y^J(s)}\td s\,,
\end{equation}
we may define the complexity of an operator as
\begin{equation}\label{defineCU1}
  \C(\U)=\min\left\{L[c]\left|~\U=c(1),~~\forall~c(s)\right.\right\}\,.
\end{equation}
It can be shown that the curve length defined in Eqs.~\eqref{defcs1} and \eqref{curLs1} gives the ``right-invariant'' Riemannian geometry.
Thus, in this framework, the essential point of a complexity theory is to define the inner product $\tilde{g}(\cdot,\cdot)$ and the complexity is the geodesic length of the curve connecting the identity and the target operator. The complexity between two states then is defined in Eq.~\eqref{cforstates1}.

Now let us review a main argument in some literatures for the reasons  why the complexity is not unitary invariant. {\it From a  quantum circuit perspective}, if the complexity of an operator $\U$ is $N$,  $\U$ may consist of
\begin{equation}
\U\approx g_Ng_{N-1}\cdots g_2g_1\,.
\end{equation}
After a unitary transformation $\U'=\O\U\O^{-1}$, we have
\begin{equation}
\U'\approx \O g_N\O^{-1}\O g_{N-1}\O^{-1}\cdots \O g_2\O^{-1}\O g_1\O^{-1}\,.
\end{equation}
If $\O g_i\O^{-1}\in\mathcal{S}$ $(i=1,2,\cdots,N)$,  the complexity of $\U'$ is still $N$. However, for a general $\O$,
\begin{equation}
\O g_i\O^{-1}\notin\mathcal{S} \,,
\end{equation}
so we have to use another fundamental gates in $\mathcal{S}$ to construct $\U'$. Thus, in general we have
\begin{equation}
\C(\U)\neq\C(\O\U\O^{-1})\,.
\end{equation}
For continuous cases, let us consider the curve length of $c(s)$ and $\O c(s)\O^{-1}$. If $H(s)$ is the generator of $c(s)$, the generator of $\O c(s)\O^{-1}$ is $\O H(s)\O^{-1}$. In general we have
\begin{equation}
\O H(s)\O^{-1}\neq H(s)\,,
\end{equation}
so the curve length is not invariant under the unitary transformation. Therefore, the complexity of an operator is also not invariant under the unitary transformation.

For  two quantum states $|\psi_1\rangle$ and $|\psi_2\rangle$, suppose that  $\U_i$ is any operator such that
\begin{equation}
|\psi_1\rangle=\U_i|\psi_2\rangle\,,
\end{equation}
and we have
\begin{equation}
\C(|\psi_1\rangle, |\psi_2\rangle)=\min\C(U_i)\,.
\end{equation}
After a unitary transformation $(|\psi_1\rangle, |\psi_2\rangle)\rightarrow(\O|\psi_1\rangle, \O|\psi_2\rangle)$,
\begin{equation}
\C(\O|\psi_1\rangle, \O|\psi_2\rangle)=\min\C(\O U_i\O^{-1})\,.
\end{equation}
 {\it From a  quantum circuit perspective}, $\C(\O U_i\O^{-1})\neq\C(\U)$ in general, so we have $\C(\O|\psi_1\rangle, \O|\psi_2\rangle)$ $\neq\C(|\psi_1\rangle, |\psi_2\rangle)$ in general.

\section{Problems of non-unitary-invariant complexity}\label{prob-non-un}
{Based on the arguments in the previous section, which seems clear {\it from a  quantum circuit perspective}, many literatures accept the conclusion that the complexity should still be non-unitary-invariant even in {\it quantum field theories}. Even though we agree that the complexity in \textit{real quantum circuits} should be non-unitary-invariant, we suspect this may not be the case in {\it quantum field theories}.
Some of problems of non-unitary-invariant complexity have been discussed in our previous papers~\cite{Yang:2018nda,Yang:2018cgx}. Here, we review some of them and also add new arguments. For simplicity, we consider only quantum mechanics but similar conclusions can be obtained in quantum field theories.}

As discussed in \cite{Yang:2018nda} and also commented by  Nielsen in this original work~\cite{Nielsen:2006:GAQ:2011686.2011688}, the Finsler (non-Riemannian) geometry is more suitable for the operators complexity. However, as i) the aim of this section is just to discuss the unitary invariance and bi-invariance and ii) the same arguments are still valid in the general Finsler geometry, we will focus on the Riemannian geometry. We use the Sachdev-Ye-Kitaev model as a concrete example to show why the Finsler (non-Riemannian) geometry is better than the widely used Riemannian geometry in our upcoming work~\cite{Yang:2019iav}.

\subsection{Left-order or right-order?} \label{sec31}
First, let us consider the product in Eq.~\eqref{timeoder1}. We may ask ``why are new operators all multiplied in the left side?''
Of course, as we noted below Eq.\eqref{condgsU}, we may use a different order.
For example, at the initial time $t_1$, we have one gate $g_1$ in the circuit, so
\begin{equation}
t=t_1,~~~\U(t_1)=g_1\,.
\end{equation}
At time $t_2>t_1$, we add a new gate at the right-side
\begin{equation}
t=t_2,~~~\U(t_2)=g_1g_2\,,
\end{equation}
and at time $t_3>t_2$
\begin{equation}
t=t_3,~~~\U(t_3)=g_1g_2g_3\,,
\end{equation}
so we can construct the operator by the right-order in time evolution
\begin{equation}\label{timeorder2}
  \U=g_1g_2\cdots g_k\,.
\end{equation}

There is no mathematical or physical reason to forbid us from constructing the operator in this `right-order' rather than the `left-order'.
This implies that, even with the same fundamental universal gates set, we have two different manners to define the complexity.
This is true also in continuous systems. For the same curve $c(s)$, we may use the order either $\mathcal{P}=\overleftarrow{\mathcal{P}}$ (left-order) or $\mathcal{P}=\overrightarrow{\mathcal{P}}$ (right-order). Following the same logic in Refs.~\cite{Nielsen1133,Nielsen:2006:GAQ:2011686.2011688,Dowling:2008:GQC:2016985.2016986,Susskind:2014jwa,Brown:2017jil,Jefferson:2017sdb} the left-order (or right-order) implies that the complexity geometry is right-invariant (or left-invariant) Riemannian geometry.

Let us denote the complexity based on the product Eq.~\eqref{timeoder1} by $\C_r$, where the subscript $r$ means `right-invariant' complexity.  Let us also denote the complexity based on the product Eq.~\eqref{timeorder2} by $\C_l$, where the subscript $l$ means `left-invariant' complexity.
If two complexities are same, the complexity geometry is bi-invariant.

It has been shown that~\cite{Yang:2018nda,Yang:2018cgx}, for a right(or left)-invariant complexity geometry, the \textit{unitary invariance implies the bi-invariance}, i.e.,
\begin{equation} \label{CrCld0}
\C_{r(l)}(\U)=\C_{r(l)}(\O\U\O^{-1})\quad \iff \quad \C_r(\U)=\C_l(\U)\,.
\end{equation}
If the complexity in quantum field theory should be only right-invariant but non-unitary-invariant, as discussed in Refs.~\cite{Susskind:2014jwa,Brown:2017jil,Jefferson:2017sdb}, two complexities $\C_r$ and $\C_l$ should be different
\begin{equation} \label{CrCld}
\C_r(\U)\neq\C_l(\U)\,.
\end{equation}

This difference will not make any  problem in real quantum circuits, since to construct a bigger circuit, we have to choose one manner (left or right) anyway and we know which manner was used in this real quantum circuit.
However, if we consider quantum field theories or other natural physical systems, we meet different situations.
For a time-evolution system, suppose that we find the time evolution operators $\U(t_i)$ and $\U(t_{i+1})$ at  time $t=t_i$ and $t_{i+1}$.
{\bf Q1:} ``How can we verify which one of the following manners
\begin{equation}
\U(t_{i+1})=g_{i+1}\U(t_i) \qquad \text{or} \qquad \U(t_{i+1})=\U(t_i)g_{i+1}' \,,
\end{equation}
is used by nature?''  The former is the left-order while the latter is the right-order.

Many literatures say the answer is the former by simply adopting the left-order without a justification. Thus, $\C_r$ is chosen to describe the complexity of quantum field systems. However, they also assume non-unitary-invariance of the complexity based on the observation from quantum circuits. This means $\C_r(\U)\neq\C_l(\U)$ as shown in Eq. \eqref{CrCld}. Now, we face the problem. It seems that there is no reason forbid nature from choosing the right-order.
If the answer to {\bf{Q1}} is the latter (the righ-order) we will have $\C_l$ as the complexity. Now, what is the correct one, $\C_r(\U)$ or $\C_l(\U)$?

One may argue that i) in all other studies in quantum mechanics/field theories, we usually use the left-order; ii) the successes in these studies are enough to show that the left-order is physically favored than the right-order.
However, this is not the case.
Suppose that $H_r(s)$ and $H_l(s)$ are the generators of the same curve $c(s)$ but in different product orders
\begin{equation}\label{defcs2}
  c(s)=\overleftarrow{\mathcal{P}}\exp\int_0^siH_r(t)\td t=\overrightarrow{\mathcal{P}}\exp\int_0^siH_l(t)\td t\,,
\end{equation}
In most case, we use $H_r(s)$ as a physical Hamiltonian to study the evolution of systems but do not consider its partner $H_l(t)$. This is simply because $H_r(s)$ and $H_l(s)$ satisfy
\begin{equation}\label{HrHl}
H_r(s)=c(s)H_l(s)c^{-1}(s)\,,
\end{equation}
i.e., $H_r(s)$ and $H_l(s)$ are different only by a unitary transformation.
So far, all observable physical properties, including thermodynamics, $n$-points correlation functions and scattering cross-sections, are equivalent under unitary transformations. Just because of this fact it is enough to consider only $H_r(s)$.

There may be another argument for allowing {\it only right/left invariant} complexity: i) we may have two different complexities $\C_{r} $ and $\C_{l}$ computed from {\it the same underlying dynamics}. ii) {though they are related by the same underlying physics, these two are just two {\it different} physical observables and do not need to be the same. iii) we may choose either one as our observable and one can be `translated' to the other.}  However, we will show that, these two complexities $\C_{r} $ and $\C_{l}$ indeed must be {\it bi-invariant} so they are the {\it same} up to an overall constant, if they are derived from {\it the same dynamics}.

\begin{figure}
  \centering
  \includegraphics[width=.6\textwidth]{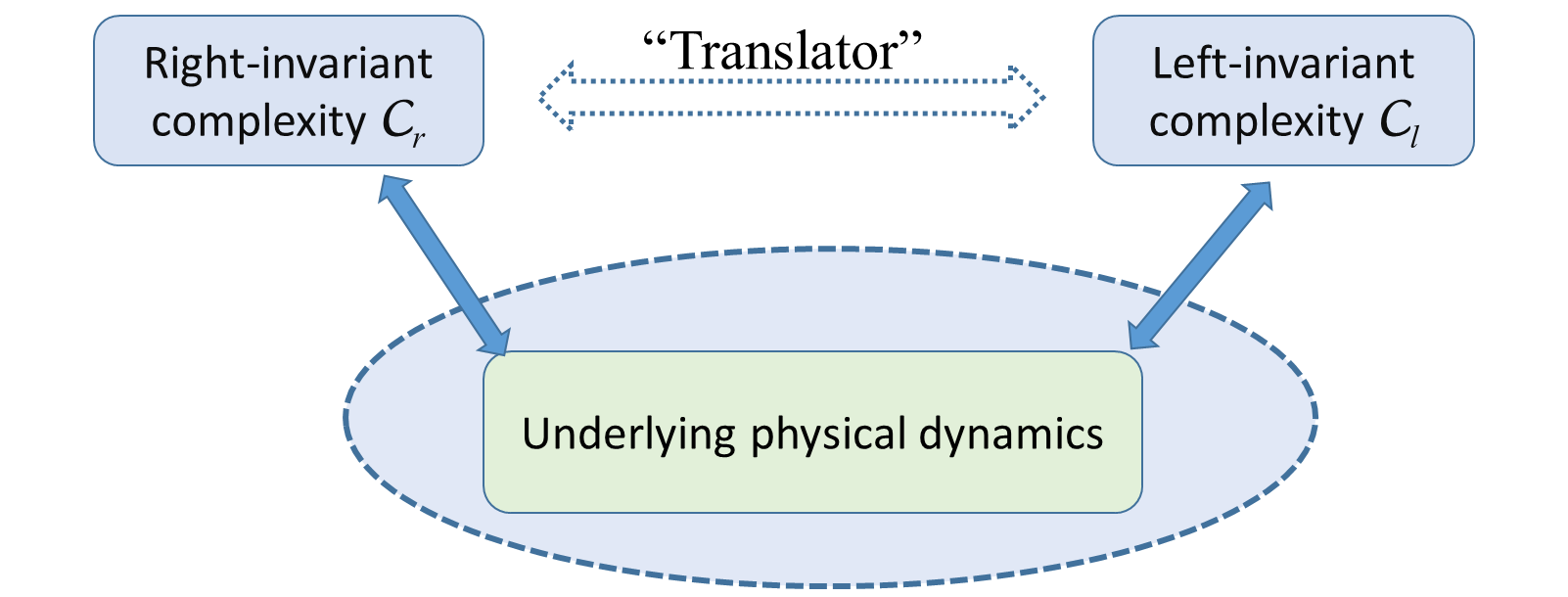}
  \caption{{As two different complexities describe the same underlying physical dynamics, there should be a ``translator'' to connect them.}} \label{trans1}
\end{figure}

Let us explain it in more detail. Suppose that we obtain the complexity in one way. After then we can use a ``translator'' $F$ to translate it into the other one (see Fig.~\ref{trans1} as a schematic explanation).
Note that two complexities do not need to have the same value. For example, for an evolution curve $c(s)$, we can compute its length $L_r[c]$ for the right invariant complexity. ($L_{r(l)}[c]$ is the $L[c]$ for the right(left)-invariant case.) The corresponding left-invariant complexity can be obtained by the ``translator'' $F$ defined as
\begin{equation}\label{convert1}
  L_l[c]=F(L_r[c])\,,
\end{equation}
which does not need to be the same as $L_r[c]$, in general. In this sense, two complexities are different but can still correspond to the same physical dynamics.
Now let us consider two curves $c_1(s)$ and $c_2(s)$. By Eq.~\eqref{convert1}, we have
\begin{equation}\label{convert2}
  L_l[c_1]=F(L_r[c_1]),~~~L_l[c_2]=F(L_r[c_2])\,.
\end{equation}
Taking the curve $c_2$ to be the right-translation of $c_1$, i.e., $c_2(s)=c_1(s)\U$, we have
\begin{equation}\label{convert3}
  L_l[c_1\U]=F(L_r[c_1\U]) = F(L_r[c_1]) =L_l[c_1]\,,
\end{equation}
where in the second equality we used $L_r$ is right-invariant. This shows that $L_l$ is also right-invariant and so bi-invariant. For the same reason, $L_r$ should be also bi-invariant. {For a semi-simple Lie group,} this means that the complexities from $L_l$  and $L_r$ are indeed the same up to an overall constant.

One may also argue that: i) the ``local'' or ``simple'' generator should produce smaller complexity while the ``non-local'' or ``hard'' generator should produce largger complexity, and ii) the ``local/simple'' generator may be transformed into ``non-local/hard'' generator under unitary transformations. These two arguments show that the complexity could be different under unitary transformations. If these are true, let us consider the time evolution in Eq.~\eqref{defcs2} and the relationship~\eqref{HrHl}. If, as what i) and ii) say, the ``local/non-local'' or ``simple/hard'' play a role in complexity and can be transformed to each other by unitary transformations, it is possible that the Hamiltonian $H_r$ may be local/simple but $H_l$  may be non-local/hard and they can be transformed. Then, a question arises: should the time evolution $c(s)$ be local/simple or non-local/hard? {We will come back to the ``locality'' again and discuss more in Sec.~\ref{commlocal}.}

These observations imply that if the complexity is non-unitary-invariant, i) $H_r(s)$ and $H_l(s)$ will give different complexities and ii) these two different complexities will correspond to different dynamics. Thus we have to choose among the left-order and the right-order and have to justify our choice: ``why is one physically more favored than the other?''
Note that if the complexity is unitary-invariant, we do not need to answer this question because $\C_r(\U) = \C_l(\U)$ as in Eq.\eqref{CrCld0}.

\subsection{Too many free parameters} \label{sec32}
It seems that the only right-invariant (non-unitary-invariant) complexity theory has too many adjustible parameters to be a controlled and predictive theory.

For example, let us consider a simple model, the harmonic oscillator of which Hamiltonian reads
\begin{equation}\label{harmH1}
  H_2=Y_1\x^2+Y_2\p^2\,.
\end{equation}
There are two generators in this Hamiltonian, $\x^2$ and $\p^2$, but they do not form a Lie algebra. To form a closed Lie algebra, we need to add the third generator $\x\p+\p\x$ to $H_2$:
\begin{equation}\label{harmH3}
  H_3=Y_1\x^2+Y_2\p^2+Y_3(\x\p+\p\x)\,.
\end{equation}
Thus, to describe the complexity geometry, we need to know 6 independent metric components in a 3-dimensional manifold, see Eq. \eqref{metricgs1}.
Until now, there is no literature which offers a way to compute metric or to  determine it by experiments.\footnote{The locality proposed in Ref.~\cite{Brown:2017jil} cannot determine the metric components, as all three generators are local generators.}
Thus, some literatures such as Refs.~\cite{Jefferson:2017sdb,Camargo:2018eof} made particular choices for metric by hand. However, the properties of the complexity in  Refs.~\cite{Jefferson:2017sdb,Camargo:2018eof} may depend on the choice of the metric, so it is not very clear how much they are general or robust.


Ref.~\cite{Camargo:2018eof} noted $\x$ and $\x\p+\p\x$ form a 2-dimensional sub-algebra
\begin{equation} \label{H2p}
H_2'=Y_1\x^2+Y_3(\x\p+\p\x) \,,
\end{equation}
and discussed the complexity for this 2-dimensional case. In this 2-dimensional case, because the complexity geometry is right (or left)-invariant, the scalar curvature is non-positive constant.  Thus, it is a hyperbolic geometry with negative constant curvature.  Because all such complexity geometries are equivalent by overall factor, the complexity geometry is completely fixed in this case.
However, it is important to note that the Hamiltonian \eqref{H2p} cannot be treated as a physical Hamiltonian because it is not bounded below, though it  can be considered as a good mathematical toy model.
Thus, for a physical model, we have to deal with \eqref{harmH3} with 6 independent metric components by which the complexity geometry is defined.  There seems no physical motivation to give large ``penalty'' to any specific metric component. If we choose the metric by hand for convenience or simplicity, the theory may not be able to capture essential and intrinsic properties of the complexity.

Note that this artificial choice will not make any  problem in real quantum circuits, since giving some artificial penalty to some gate is a part of the game.
We also emphasize that the `bounded-below' of the Hamiltonian is a very important property when we consider the complexity in quantum mechanics/field theories. On the contrary, in quantum circuits, we never need to worry if a `Hamiltonian', in the sense of $H$ in Eq.\eqref{condgsU}, is bounded below.

{If we consider a more realistic Hamiltonian $H=\vec{p}^2/(2m)+V(\vec{x})$, the problem of ``too many free parameters'' becomes more serious.
For example, with the ``central potential'' $V(\vec{x})=1/|\vec{x}|^2$, the Hamiltonian contains two generators $\vec{p}^2$ and $V(\vec{x})$. To build the complexity geometry, we have to add their commutators into the generators in order to form a Lie algebra. Firstly, we have to add $e_3:=[\vec{p^2},V(\vec{x})]$ into the bases because
$$[\vec{p}^2,V(\vec{x})]=-i(\vec{p}\cdot\vec{\nabla}V(\vec{x})+\vec{\nabla}V(\vec{x})\cdot\vec{p})\neq0\,.$$
However, one can check that $e_4:=[e_3,\vec{p}^2]\neq0$ and $e_5:=[e_3,V(\vec{x})]\neq0$. As a result, we need  to add also $e_4$ and $e_5$ into the bases, but it turns out that $[e_3,e_4], [e_3, e_5], [e_4,e_5]$ are all nonzero. Thus we need to add more and more generators into the bases. Finally, to form a Lie algebra we have to introduce infinitely many generators. Following the idea of Refs.~\cite{Nielsen1133,Nielsen:2006:GAQ:2011686.2011688,Dowling:2008:GQC:2016985.2016986,Susskind:2014jwa,Brown:2017jil,Chapman:2017rqy,Khan:2018rzm,Magan:2018nmu,Chapman:2018hou,Camargo:2018eof} we need to define infinite penalties by hand for these generators. For a different potential $V(\vec{x})$, we have to do that case by case.}


\subsection{Conflict with the framework of quantum mechanics/field theories}

Strictly speaking, the aforementioned two issues in subsections \ref{sec31} and \ref{sec32} only show unnatural aspects of non-unitary-invariant complexity. However, in this subsection, we suspect that it is even possible that there is an inconsistency between the non-unitary-invariant complexity and the current framework of quantum mechanics/field theory.

Let us again consider the harmonic oscillator, with the Lagrangian.
\begin{equation}\label{HVx1}
  L(x,\dot{x},t)=\frac{\dot{x}^2}{2m}-\frac{k}2x^2\,,
\end{equation}
where $m$ and $k$ are two positive constants.
By the Lagendre transformation we  obtain the Hamiltonian
\begin{equation}\label{defL2H1}
  H(x,p,t)=\frac{p^2}{2m}+\frac{k}2x^2,
\end{equation}

The physics will not be changed if we add a total divergence term into the Lagrangian
\begin{equation}
L(x,\dot{x},t)\rightarrow \tilde{L}(x,\dot{x},t)=L(x,\dot{x},t)+\phi'(x)\dot{x}\,.
\end{equation}
%
%
Accordingly, the Hamiltonian for $\tilde{L}(x,\dot{x},t)$ reads
\begin{equation}\label{defL2H1}
  \tilde{H}(x,p,t)=\dot{x}p-\tilde{L}(x,\dot{x},t)=\frac{(p-\phi')^2}{2m}+\frac{k}2x^2=H(x,p-\phi',t)\,,
\end{equation}
where
\begin{equation}
p=\frac{\partial\tilde{L}}{\partial\dot{x}}=\frac{\dot{x}}{m}+\phi'(x)\,.
\end{equation}
Though two Hamiltonians $\tilde{H}(x,p,t)$ and $H(x,p,t)$ look different, we know they have the equivalent physics. This suggests that the complexities given by $\tilde{H}(x,p,t)$ and $H(x,p,t)$ should be the same, i.e.,
\begin{equation}\label{restricgs1}
  \tilde{g}(\tilde{H},\tilde{H})=\tilde{g}(H,H)\,.
\end{equation}
If we note that
\begin{equation}\label{un1tary1}
   \tilde{H}(x,p,t)=H(x,p-\phi',t)=e^{i\phi(x)}H(x,p,t)e^{-i\phi(x)}
\end{equation}
we will obtain a very important symmetry for the complexity geometry
\begin{equation}\label{restricgs3}
 \forall\phi(x),~~~ \tilde{g}(e^{i\phi(x)}He^{-i\phi(x)},e^{i\phi(x)}He^{-i\phi(x)})=\tilde{g}(H,H)\,.
\end{equation}
{The same result can be obtained by an easier manner with the help of U(1) gauge symmetry. A neutral Hamiltonian $H(\vec{x},\vec{p};t)$ and a charged Hamiltonian $H(\vec{x},\vec{p}-q\vec{A}(\vec{x});t)-q\Phi(\vec{x})$ cannot be distinguished if $\Phi = \vec{A} =0$. The $U(1)$ gauge symmetry implies Hamiltonians $H(\vec{x},\vec{p}-q\vec{A}(\vec{x});t)-q\Phi(\vec{x})$ and $H(\vec{x},\vec{p}-q\vec{A}(\vec{x})+\vec{\nabla}\phi;t)-q\Phi(\vec{x})$ cannot be  distinguished in all cases.  Taking $\Phi=\vec{A}=0$, we conclude the physics of $H(\vec{x},\vec{p};t)$ and $H(\vec{x},\vec{p}-\vec{\nabla}\phi;t)$ are not  distinguished. Restricting it into one-dimensional case, we find that physics of $H(x,p-\phi',t)$ and $H(x,p,t)$ are indistinguishable, so Eq.~\eqref{restricgs3} follows.}

The above arguments can be generalized to arbitrary Lagrangian systems. The symmetry~\eqref{restricgs3} is a fundamental symmetry for the complexity of all Lagrangian systems.

In fact, in addition to the above example, we can find infinitely many similar examples. As another simple example let us consider two Hamiltonians
\begin{equation}\label{Hagroup1}
  H_1=\frac{\p^2}{2m}+\frac{k}2\x^2,~~~~H_1'=H_1+ak\x+\frac{ka^2}2\,.
\end{equation}
%
From a viewpoint of quantum circuits, as $ak\x+ka^2/2$ is a nontrivial operator, $H_1$ and $H_1'$ need to be simulated by different quantum circuits so there is no reason to expect that the complexity of the operators generated by them are same. However, note that
\begin{equation}
H_1'=H_1+ak\x+\frac{ka^2}2=\frac{\p^2}{2m}+\frac{k}2(\x+a)^2\,,
\end{equation}
which means that $H_1'$ is obtained just by constant-shifting the coordinates of $H_1$ so both are equivalent. Thus we may well expect
\begin{equation}
\tilde{g}(H_1,H_1)=\tilde{g}(H_1',H_1')\,.
\end{equation}
This is valid also for more general cases, i.e., for an arbitrary potential $V(\x)$, the following two Hamiltonians{
\begin{equation}\label{Hagroup1}
  H_1=\frac{p^2}{2m}+V(\x),~~~~H_1'=\frac{p^2}{2m}+V(\x)+\sum_{n=1}^\infty \frac{V^{(n)}(\x)}{n!}a^n\,,
\end{equation}
should give the same complexity.  Here $V^{(n)}(x):=\td^n V(x)/\td x^n$. For general $V(x)$, the additional terms in $H_1'$ may be very complicated, and, from the perspective of quantum circuits, may need many additional gates to realize. However, we know $H_1'=\frac{p^2}{2m}+V(\x+a)$ and $H_1$ describe the equivalent physics, because $H_1'$ is obtained by just shifting the origin in the coordinate $x$.}
Because $H_1'=e^{-ia\p}H_1e^{ia\p}$ we have another general unitary symmetry for the complexity geometry
\begin{equation}\label{tranla1}
  \forall a\in\mathbb{R},~~~\tilde{g}(H_1,H_1)=\tilde{g}(e^{-ia\p} H_1e^{ia\p},e^{-ia\p}H_1e^{ia\p}) \,.
\end{equation}

In addition to the constant shift of the coordinates, we can also rescale the coordinates for the harmonic oscillator Hamiltonian~\eqref{defL2H1},
\begin{equation}\label{recalx}
x\rightarrow\xi x\,,
\end{equation}
which induces a transformation of momentum
\begin{equation}\label{recalp}
p\rightarrow\xi^{-1}p\,.
\end{equation}
%
The Hamiltonian becomes
\begin{equation}\label{Hxis1}
  H_{\xi}(x,p,t)=H(\xi x,\xi^{-1}p,t)=e^{iW(\xi)}H(x,p,t)e^{-iW(\xi)}=\frac{p^2}{2m\xi^2}+\frac{\xi^2k}2x^2\,,
\end{equation}
with
\begin{equation}
W(\xi)=-\frac{i}2(xp+px)\ln\xi\,.
\end{equation}
As the only relevant parameter in the Hamiltonian~\eqref{defL2H1} is the frequency $\omega=\sqrt{k/m}$ and $\omega$ is invariant under the transformations~\eqref{recalx} and \eqref{recalp}, we find that the complexity geometry has another symmetry for Hamiltonian~\eqref{defL2H1}
\begin{equation}\label{restricgs4}
 \forall\xi>0,~~~ \tilde{g}(e^{iW(\xi)}He^{-iW(\xi)},e^{iW(\xi)}He^{-iW(\xi)})=\tilde{g}(H,H)\,.
\end{equation}

In summary, we find that  Eqs.~\eqref{restricgs3}, \eqref{tranla1} and \eqref{restricgs4}, straightforwardly derived from the fundamental formalism of quantum mechanics/field theory, all suggest that the complexity need to be unitary-invariant. In other words, non-unitary-invariant complexity may not be compatible with the fundamental principles of the quantum field theory.\footnote{For another argument supporting unitary-invariant complexity see  section~4.1.2 of Ref.~\cite{Yang:2018tpo}.}
Here, we want to again emphasize that such conflicts will not appear in real quantum circuits, as the real quantum circuits are not based on the Lagrangian formalism.

The above conclusion can be understood from a more general perspective of quantum field theory.  The generating functional $Z[J]:=\Tr(e^{- i H[J] t })$ plays a central role in the current frameworks of quantum field theory.  It is assumed that all physical observables can be obtained from the generating functional, which has been confirmed in experiments and observations, from laboratories to cosmos. Even the AdS/CFT correspondence is expressed in terms of the equivalence of  the generating functional. Two Hamiltonians $H[J]$ and $\tilde{H}[J]=\U^\dagger H[J]\U$ have the same generating functional, so they are expected to have the same physics. This means $H[J]$ and $\U^\dagger H[J]\U$ should give the same complexity, which is equivalent to the bi-invariance of the complexity. The generating functional does not play such a central role in real quantum circuits, so the complexity there does not need to be bi-invariant or unitary invariant.

Let us recall the issue in the previous subsection: too many free parameters. Indeed, this issue can be resolved if we adopt the unitary-invariance of the complexity, which can give a constraint to the complexity geometry.
For example, let us apply the symmetries~\eqref{restricgs3} and \eqref{restricgs4} to the Hamiltonian~\eqref{harmH3} and specify $\phi(x)=x^2/2$. It yields the components of metric $\tilde{g}_{IJ}$ in the bases $\{\x^2,\p^2,\x\p+\p\x\}$\footnote{See Ref. \cite{Yang:2018cgx} for more details and other independent arguments supporting it.}
\begin{equation} \label{metric000}
\tilde{g}_{11}=\tilde{g}_{31}=\tilde{g}_{21}=\tilde{g}_{22}=0,~~~\tilde{g}_{33}=-2\tilde{g}_{12}\,.
\end{equation}
Apart from one overall factor all 6 components of the metric are fixed so we do not have any freedom to choose the so-called penalties by hand. In other words, the penalties are given by theory itself. Thanks to this we may study intrinsic property of the complexity of the Hamiltonian~\eqref{harmH3} without any artificial choice, as discussed in \cite{Yang:2018tpo}.

\subsection{A short summary and comments} \label{shortsummary1}
Let us make a short summary on what problems may arise if we {\it assume} that the complexity in quantum mechanics/field theory is non-unitary-invariant.

\begin{enumerate}
\item[(1)] For a given physical situation, there are two ways to define the complexities $\C_r$ and $\C_l$. If the complexity is non-unitary-invariant, in general, $\C_r$ and $\C_l$ may be different and can not tell us the same physics, but there is no good physical reason to tell which one is correct.
\item[(2)] If the complexity is non-unitary-invariant, there are too many free parameters in the theory. Along this line, current studies are based on some artificial choices of the parameters, which make intrinsic understanding of the complexity difficult.
\item[(3)] The non-unitary-invariant complexity may be in general in conflict with the fundamental method and symmetry of quantum physics based on the Lagrangian/Hamiltonian formalism because they suggest the complexity is unitary-invariant.
\item[(4)] The framework of the quantum field theory assumes that physical observables are encoded in the generating functional. However, the non-unitary-invariant complexity implies that the same generating functional can give different physics.
\end{enumerate}

All these four problems do not arise in the complexity of quantum circuits. They arise if we simply adopt the ``non-unitary-invariance'' of the complexity for quantum field theory. Many literatures, such as Refs.~\cite{Brown:2017jil,Jefferson:2017sdb,Yang:2017nfn,Chapman:2017rqy,Khan:2018rzm,Magan:2018nmu,Chapman:2018hou,Camargo:2018eof}, favor the non-unitary-invariant complexity and select penalties to discuss their physics. This has no problem in quantum circuits as real circuits are made by human and human has the right to define the penalty of every gate.
However, in quantum field theory constructed by nature, it will be more desirable if the penalties can be given by more fundamental theories or measured by experiments. Unitary-invariance may play a role in this respect by giving some constraints, see for example Ref.~\cite{Yang:2018tpo,Yang:2018cgx}.

\section{Comment on locality: apparent locality vs intrinsic locality}\label{commlocal}
It has been argued in Ref.~\cite{Brown:2017jil} that the complexity has something to do with ``locality''. The concept of ``locality'' will be more clarified later, but for now, we note, roughly speaking, ``local" theory is ``simple'' and ``non-local'' theory is ``complex''.\footnote{{In our opinion,  this local/simple and non-local/complex relation may not be so robust. In principle, it is possible to have ``less complex'' non-local operator than a simple operator. Therefore, ``more non-local'' and ``more complex'' may not have a strong relationship in general.}} However, the unitary transformation in general seems to change the ``locality'' of the theory so should change the complexity. Therefore, one may conclude the complexity is non-unitary invariant.

Based on this argument, many literatures have tried to deal with non-unitary invariant or non-bi-invariant complexity by choosing some parameters in their theory by hand. In this section, we want to show that
\begin{itemize}
\item There are two kinds of locality, the ``apparent locality'' and ``intrinsic locality" (we will present detailed definitions later). The apparent locality may vary under the unitary transformation but intrinsic locality will not.  
    We think the ``locality'' used in Ref.~\cite{Brown:2017jil}  is an ``apparent locality'.
\item  The apparent locality, though is useful in some cases, it cannot grasp the essential differences of local theory and non-local theory {regarding the complexity.} For example, suppose that the ``apparently'' local theory $H_l$ becomes the ``apparently'' non-local theory $H_r$ by a unitary transformation \eqref{defcs2}. In this case, how do we know if the evolution operator $c(s)$ in Eqs.~\eqref{defcs2} stands for a local theory or non-local theory? The logical answer to avoid contradiction in Eqs.~\eqref{defcs2} would be it corresponds to an ``intrinsically'' local theory.
\end{itemize}
%


Let us now go into more details.
We start with clarifying the meaning of  ``locality''\footnote{The ``locality'' can have different meanings in other contexts.
First, in the context of the quantum states, it means that the corresponding wave functions is well localized, i.e., $\psi(x)\rightarrow0$ rapidly if $x\rightarrow\pm\infty$. In the context of the field operator $\phi(x)$ it has something to do with local commutativity or microscopic causality, i.e., two fields are space-like separated and the fields either commute or anticommute.} used in Ref.~\cite{Brown:2017jil}.
It means that the mathematical expressions of Hamiltonian or Lagrangian  contain only local interactions and finitely many derivatives. We will call it ``apparent locality'' or an ``apparently local theory''.  A theory will be called ``apparently non-local'' if it is not an apparently local theory.

For example, the following Lagrangian is apparently local
\begin{equation}
L_1=(\partial_t\phi(x,t))^2-W(\partial_x\phi(x,t))-V(\phi(x,t))\,,
\end{equation}
where $W$ and $V$ are arbitrary two smooth functions. The following three Lagrangians are apparently non-local
\begin{equation}\label{defLs2}
  L_{2}=(\partial_t\phi(x,t))^2-W(\partial_x\phi(x,t))-\phi(x,t)\phi(x+a,t)\,, \quad a\neq0\,,
\end{equation}
\begin{equation}\label{defLs3}
  L_3=(\partial_t\phi(x,t))^2-W(\partial_x\phi(x,t))-\int\td y\phi(x,t)\phi(x+y,t)\,,
\end{equation}
and
\begin{equation}\label{defLs4}
  L_4=(\partial_t\phi(x,t))^2-V(\phi(x,t))-\phi(x,t)\left(\sum_{n=0}^{\infty}\frac{a^n}{n!}\partial_x^n\phi(x,t)\right)\,, \quad a\neq0\,.
\end{equation}
The $L_4$ is apparently non-local theory as
$$\sum_{n=0}^{\infty}\frac{a^n}{n!}\partial_x^n\phi(x,t)=\phi(x+a,t)\,.$$

In general, we can also defined ``apparent $k$-locality'' and an ``apparently $k$-local theory'', in which the Hamiltonian and Lagrangian contains interactions involving $k$ different points.
For example, $L_1$ is apparently 1-local, while $L_2, L_3$ and $L_4$ are all apparently 2-local.
The Sachdev-Ye-Kitaev model is a quantum-mechanical system comprised of $N$ (an even integer) Majorana fermions $\chi_i$ with the Hamiltonian
\begin{equation}\label{defsykH1}
  H_{\mathrm{SYK}}=\sum_{i<j<k<l}^{N}J_{ijkl}\chi_i\chi_j\chi_k\chi_l\,,
\end{equation}
where the coefficients $J_{ijkl}$ are drawn at random from a Gaussian distribution. This is apparently 4-local as it involves the interactions of four different points.

To explain why the apparent locality used in Ref.~\cite{Brown:2017jil} may not be intrinsic and depends on our (human's) preferences, let us consider a similar example in general relativity. We may ask if the following metric
\begin{equation}\label{metric1}
  \td s^2=-(1-e^{2x})\td t^2+2(te^{2x}-1)\td t\td x+(t^2e^{2x}-1)\td x^2 \,,
\end{equation}
describes a flat spacetime or not? Naively (or ``apparently'' in our terminology), the metric looks not flat because it is different from $\td s^2=-\td t^2+\td x^2$. However, after the following coordinates transformation
\begin{equation}
\tau=t+x,~~\xi=t e^{x}\,,
\end{equation}
the above metric becomes $\td s^2=-\td \tau^2+\td\xi^2$, which is indeed flat. As is well known, flatness cannot be easily understood simply by looking at the ``apparent'' form of metric components.

A similar reasoning may apply to ``locality."
Let us now ask if the following Lagrangian
\begin{equation}\label{defLeg1}
  L=\left[\int\td yh(x,y)\partial_t\phi(y,t)\right]^2-\left[\partial_x\int\td yh(x,y)\partial_y\phi(y,t)\right]^2-\left[\int\td yh(x,y)\phi(y,t)\right]^2 \,,
\end{equation}
is ``local'' or not.  Here the integration range is $-\infty<x<\infty$, the function $h(x,y)$ satisfies
\begin{equation}\label{prophxy}
  \partial_xh(x,y)=-\partial_yh(x,y),~~h(x,y)|_{x\rightarrow\pm\infty}=h(x,y)|_{y\rightarrow\pm\infty}=0\,.
\end{equation}
and there is a function $\tilde{h}(x,y)$ such that
\begin{equation}\label{prophxy2}
  \int\td xh(x,y_1)\tilde{h}(x,y_2)=\delta(y_1-y_2),~~\int\td xh(x_1,y)\tilde{h}(x_2,y)=\delta(x_1-x_2)\,.
\end{equation}
This theory is ``apparently non-local'' as it involves the interactions of different points. However, making a variable transformation
\begin{equation}\label{defpsiohi1}
  \psi(x,t)=\int h(x,y)\phi(y,t)\td y\,,
\end{equation}
and noting the fact
\begin{equation}
  \int\td yh(x,y)\partial_y\phi(y,t)=h(x,y)\phi(y,t)|_{y=-\infty}^{y=\infty}-\int\td y\partial_yh(x,y)\phi(y,t)=\int\td y\partial_xh(x,y)\phi(y,t)\,,
\end{equation}
we have
\begin{equation}\label{defLeg2}
\begin{split}
  L&=\left[\partial_t\int\td yh(x,y)\phi(y,t)\right]^2-\left\{\partial_x\int\td y[\partial_xh(x,y)]\phi(y,t)\right\}^2-\left[\int\td yh(x,y)\phi(y,t)\right]^2\\
  &=[\partial_t\psi(x,t)]^2-[\partial_x^2\psi(x,t)]^2-\psi(x,t)^2\,.
  \end{split}
\end{equation}
After a suitable variable transformation, we find that the new Lagrangian~\eqref{defLeg2} becomes ``apparently local''.

To be self-consistent, it is necessary to check that if the variable transformation can keep the canonical commutation (or anticommutation) relation or not. The canonical momentum of $\phi$ for the Lagrangian~\eqref{defLeg1} reads
\begin{equation}\label{defpiphi}
  \pi_\phi(x,t):=\frac{\delta L}{\delta\partial_t\phi(x,t)}=2h(x,y)\int\td zh(y,z)\partial_t\phi(z,t)\,.
\end{equation}
We see that the momentum depends on the value of $\partial_t\phi$ in the whole space. The quantization can be achieved by imposing the following canonical commutation (or anticommutation) relation
\begin{equation}\label{commutator1}
  [\phi(x_1,t),\pi_\phi(x_2,t)]=i\delta(x_1-x_2)\,.
\end{equation}
From the Lagrangian~\eqref{defLeg2} we can obtain canonical momentum of $\psi(x,t)$
\begin{equation}\label{defpipsi}
  \pi_\psi(x,t):=\frac{\partial L}{\partial\partial_t\psi(x,t)}=2\partial_t\psi(x,t)=2\int h(x,y)\partial_t\phi(y,t)\td y\,.
\end{equation}
Combining the orthogonal relationship~\eqref{prophxy2} and the relationship~\eqref{defpiphi}, we obtain
\begin{equation}
\pi_\psi(x,t)=\int\tilde{h}(x,y)\pi_{\phi}(y,t)\td y\,.
\end{equation}
We see that, under the variable transformation~\eqref{defpsiohi1}, the canonical momentum is transformed as
$$\pi_\phi\rightarrow\pi_{\psi}=\int\tilde{h}(x,y)\pi_{\phi}(y,t)\td y\,.$$
Then we can check the new variables $\psi$ and $\pi_\psi$ satisfy the same canonical commutation (or anticommutation) relation
\begin{equation}\label{commutator2}
\begin{split}
  [\psi(x_1,t),\pi_\psi(x_2,t)]&=\left[\int h(x_1,y_1)\phi(y_1,t)\td y_1,\int \tilde{h}(x_2,y_2)\pi_{\phi}(y_2,t)\td y_2\right]\\
  &=\int h(x_1,y_1)\tilde{h}(x_2,y_2)\td y_1\td y_2[\phi(y_1,t),\pi_{\phi}(y_2,t)]\\
  &=i\int h(x_1,y_1)\tilde{h}(x_2,y_2)\td y_1\td y_2\delta(y_1-y_2)\\
  &=i\int h(x_1,y_1)\tilde{h}(x_2,y_1)\td y_1=i\delta(x_1-x_2)\,.
  \end{split}
\end{equation}
Checking such a self-consistence is necessary as not all variable transformations keep the canonical commutation (or anticommutation) relation. If a variable transformation changes these canonical relations, it will change physics.

We have found that, by a suitable variable transformation, an apparently non-local theory~\eqref{defLeg1} can be changed into an apparently local theory~\eqref{defLeg2}. One may argue that, though in term of $\psi(x)$, the Lagrangian~\eqref{defLeg2} has a local form, the field $\psi(x)$ contains integration of $\phi(x)$ and Eq.~\eqref{defLeg2} should still be treated as a non-local theory. About this argument, we would like to point out that Eqs.~\eqref{defpsiohi1} and \eqref{prophxy2} imply
\begin{equation}\label{defpsiohi2}
  \phi(x,t)=\int \tilde{h}(x,y)\psi(y,t)\td y,~~\psi(x,t)=\int h(x,y)\phi(y,t)\td y \,.
\end{equation}
The field $\phi(x)$ is also the integration of field $\psi(x)$ so there is no  reason to say that only $\phi(x)$ can be treated as a physical field operator but $\psi(x)$ can not be. After we choose $\psi(x)$ as the field operator, the theory becomes apparently local. Or we can say that, the Lagrangian~\eqref{defLeg1} is apparently non-local because we choose a ``bad'' field operator.

Some apparently non-local theories can be transformed into apparently local theories by suitable variables transformations, but some apparently non-local theories can not. For example, the three Lagrangians defined in Eqs.~\eqref{defLs2}, \eqref{defLs3} and \eqref{defLs4} can not be written in terms of apparently local Lagrangians by variable transformations. This means that, though Lagrangian \eqref{defLeg1} and Langrangians \eqref{defLs2}~\eqref{defLs3}~\eqref{defLs4} are all apparently non-local, they have essential differences. On the other hand, all apparently local theories can be transformed into the apparently non-local theories by suitable variable transformations. There are some freedoms in choosing the field operators and making variables transformations, and the apparent locality depends on the choices of variables and variable transformations. 
The Lagrangian~\eqref{defLeg1} looks like non-local because we choose the ``bad'' field operator rather than the theory is really non-local, which is similar to the aforementioned metric example: the metric~\eqref{metric1} ``apparently'' (naively) looks like curved spacetime because we choose ``bad'' coordinates rather than the spacetime is really curved.

If we want the concept of locality to be defined by some intrinsic properties of physical theories it should be defined as a way which does not depends on any specific choice of field operator. It is similar to general relativity: the flatness should be defined by a manner which does not depend on any specific choice of coordinates. Thus, for a field theory, it is more useful to define an ``intrinsic locality'' in such way:
\begin{enumerate}
\item[]\textit{If there is one suitable variable transformation to transform a Lagrangian into an ``apparently local'' form keeping the canonical commutation (or anticommutation) relation, then the theory is intrinsically local; if such a variable transformation does not exist, then the theory is intrinsically non-local. }
\end{enumerate}
The intrinsic locality will not be changed by variable transformations.

The above definition gives us a way to verify if a theory is intrinsically local or not. However, it is difficult to verify the existence of such a variable transformation for a general complicated Lagrangian.
In general relativity, it is also difficult to verify if there is a coordinates transformation so that the metric components becomes the  Minkowski form. However, the Riemann tensor offers us a powerful tool to judge the flatness even if we do not know such coordinates transformation.
Do we have any method to verify the intrinsic locality for arbitrary given Lagrangian even if we do not know the corresponding variable transformation? This question seems very interesting in both mathematics and physics. We do not have a complete answer. However, here we would like to present a simple relevant proposition:
\begin{enumerate}
\item[]\textit{A given Lagrangian $L$ in term of a field operator $\phi$ describes an intrinsically local theory if and only if its generating functional $Z[J]$ equals to the generating functional of an apparently local theory.}
\end{enumerate}
It can be partly justified by the \textit{Wightman reconstruction theorem}.\footnote{Strictly speaking the Wightman reconstruction theorem is valid for free scalar and spinor theories.} 
Because $Z[J]$ is the same as the generating functional of an apparently local theory,  its all $n$-point functions are the same as the $n$-point functions of an apparently local theory. The \textit{Wightman reconstruction theorem} says that such two theories are different only up to a unitary transformation. This means that we can find a unitary transformation $\phi\rightarrow\psi=\hat{U}\phi\hat{U}^\dagger$, under which the Lagrangian $L$ becomes apparently local and  the canonical momentum transforms as $\pi_{\phi}\rightarrow\pi_{\psi}=\hat{U}\pi_{\phi}\hat{U}^\dagger$. Such a unitary transformation is just a linear transformation and keep the canonical commutation (or anticommutation) relation unchanged. Thus, the proposition follows. This proposition shows that the intrinsic locality is also encoded in the generating functional. As a direct corollary, we have a conclusion:
\begin{enumerate}
\item[]\textit{If a Hamiltonian $H$ describes an intrinsically local (non-local) theory, then its arbitrary unitary transformation $H\rightarrow\U H\U^\dagger$ still describes an intrinsically local (non-local) theory. }
\end{enumerate}
We see that, the intrinsic locality, like the unitary-invariant complexity, is the unitary invariant quantity of a theory.

In general relativity, we know that the information of flatness is encoded in the Riemann curvature tensor. We have also found that the intrinsic locality is encoded in the generating functional. Then what is the essential property of the generating functional for an intrinsically local theory? We think this is an interesting question to be discussed more.

To conclude, we argue in this section that the locality discussed in many literatures such as Ref.~\cite{Brown:2017jil} may be a kind of ``apparent locality'', which depends on one's choice of field operator (or ``coordinate'') so may not be able to grasp the essential differences between the local theory and non-local theory. The locality should be defined in an intrinsic way. If the generating functional of a theory is the same as an apparently local theory, then the theory is intrinsically local. {Such an intrinsic locality is unitary invariant and is consistent with the unitary invariant complexity. }

\section{Unitary-invariant complexity of quantum states}\label{C-states}

From the discussions in the above section, we find  that, although the {\it unitary-invariant} complexity may not be suitable for real quantum circuits, it is a natural (or the only) candidate for the complexity of quantum states in quantum mechanics/field theory. In this section we propose how to construct the unitary-invariant complexity formula between states. We show our proposal is compatible with previous research such as path-integral complexity/the Liouville action, holographic CV/CA conjecture. Our proposal is not only compatible with them but also clarify their unresolved issues such as the identification of the reference state of the holographic conjectures.

\subsection{How to construct the unitary-invariant complexity}
{The complexity between two quantum states in quantum circuits is usually defined by the minimal complexity of the operators which can transform one to the other,
\begin{equation}\label{cforstates0}
  \C(|\psi_1\rangle,|\psi_2\rangle)=\min\{\C(\U)|~\forall \U, s.t., |\psi_1\rangle=\U|\psi_2\rangle\}\,.
\end{equation}
As the complexity is a dimensionless quantity, in principle, we can define a deformed complexity such that
\begin{equation}\label{cforstates1}
  \bar{\C}(|\psi_1\rangle,|\psi_2\rangle)=\min\{\bar{f}(\C(\U))|~\forall \U, s.t., |\psi_1\rangle=\U|\psi_2\rangle\}\,.
\end{equation}
Here $\bar{f}(x)$ is a monotonically increasing function and satisfies $\bar{f}(x)\geq0$ and $\bar{f}(0)=0$. This deformation does not lose any physical information of the complexity but can bring many conveniences. We will see later that, by introducing such deformation, we can connect the path-integral complexity, the CA and CV conjectures together. The deformation can be determined uniquely by the aid of the holographic conjectures. In the following part of this paper, when we talk about the complexity between states, it means this deformed complexity.
}

As $\bar{\C}(|\psi_1\rangle, |\psi_2\rangle)$ is invariant under the unitary transformation and unitary invariants formed by two quantum states $|\psi_1\rangle$ and $|\psi_2\rangle$ can be only a function of their inner product, the complexity will be of the form
\begin{equation}\label{cforinner1}
  \bar{\C}(|\psi_1\rangle, |\psi_2\rangle)=\min\{\bar{f}(\C(\U))|~\forall \U, s.t., |\psi_1\rangle=\U|\psi_2\rangle\}=f(|\langle\psi_1|\psi_2\rangle|)\,,
\end{equation}
where we have to determine the functional form $f(x)$. {We will show later that this function can be determined uniquely by a result from the holographic conjectures.}

Before we explain how to determine $f(x)$, let us first make a few comments on Eq.~\eqref{cforinner1}.
The formula~\eqref{cforinner1} looks too simple and naive:  there is no freedom to choose ``fundamental gates'' or ``penalties''. Thus, from a viewpoint of quantum circuits, this formula cannot be correct.
Our purpose is not to propose a new complexity theory for quantum circuits but to find a suitable definition of the complexity in quantum mechanics/field theory, where the volume and degrees of freedom are both infinite.
We will show in the following subsections that, though our proposal has a simple expression, its physical contents are not simple at all. In Sec.~\ref{answers1}, we will further address a few issues about this formula.

Let us now explain how to determine the function $f(x)$ in Eq.~\eqref{cforinner1}.
One useful guide is to consider the holographic results, e.g., the CV or CA conjectures~\cite{Stanford:2014jda,Brown:2015bva}, where it was noted that the complexity between a boundary state and an unknown reference state is proportional to the volume at the boundary time slices when the volume is large enough and the boundary state is uniform, i.e.,
\begin{equation}\label{CproptoV1}
  \bar{\C}(|\psi_1\rangle,|\psi_2\rangle)\propto V_{\text{bd}},~~~\text{if}~V_{\text{bd}}\rightarrow\infty\,.
\end{equation}
Here, $V_{\text{bd}}$ is the {volume of boundary states}, NOT the volume in any bulk region. This is very different from the holographic entanglement entropy.

\begin{figure}
  \centering
  \includegraphics[width=.4\textwidth]{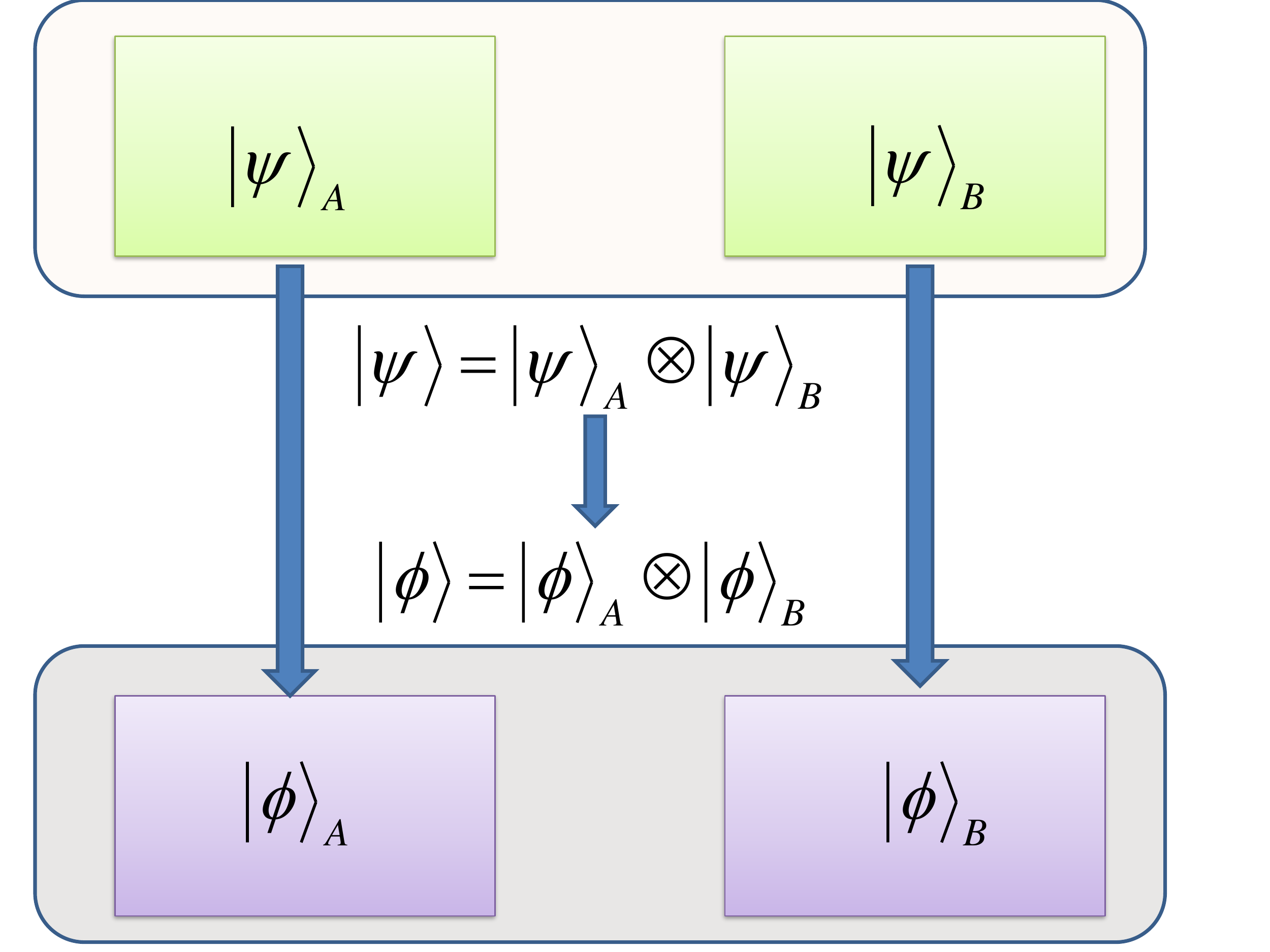}
  \caption{Schematic diagram for the complexity between two states which contain two independent sub-systems $A$ and $B$, see Eq.~\eqref{subdensity1}.  } \label{subAB}
\end{figure}

In order to see what we can obtain from this holographic property, let us consider the complexity between two states $|\psi\rangle$ and $|\phi\rangle$ which contain two independent sub-systems $A$ and $B$, e.g. see the Fig.~\ref{subAB}.
The systems $A$ and $B$ are locally the same and have the volume $V_A$ and $V_B$, respectively. When two sub-systems are separated far enough, the reference ($|\psi\rangle$) and target  ($|\phi\rangle$) state of $A\cup B$ can be written in terms of the direct product of these two independent sub-systems approximately
\begin{equation}\label{subdensity1}
  |\psi\rangle=|\psi\rangle_A\otimes|\psi\rangle_B\,, \qquad |\phi\rangle=|\phi\rangle_A\otimes|\phi\rangle_B\,.
\end{equation}

By the holographic result \eqref{CproptoV1}, we have the following relationships
\begin{equation}\label{relCVABs1}
  \bar{\C}(|\psi\rangle_A,|\phi\rangle_A)=cV_A\,, \quad \bar{\C}(|\psi\rangle_B,|\phi\rangle_B)=cV_B\,, \quad \text{and} \quad \bar{\C}(|\psi\rangle,|\phi\rangle)=c(V_A+V_B)\,,
\end{equation}
with a factor $c$. From Eqs.~\eqref{relCVABs1} and~\eqref{subdensity1} we obtain the following property.
%
%
%
\begin{enumerate}
\item [] {\bf Extensive property}: the complexity of the product states of continuous systems in thermodynamic limit is extensive
    i.e.,
\begin{equation}\label{relCVABs2b}
  \bar{\C}(|\psi\rangle_A\otimes|\psi\rangle_B, |\phi\rangle_A\otimes|\phi\rangle_B)=\bar{\C}(|\psi\rangle_A,|\phi\rangle_A)+\bar{\C}(|\psi\rangle_B,|\phi\rangle_B)\,,
\end{equation}
if the states $|\psi\rangle_A, |\psi\rangle_B, |\phi\rangle_A$ and $|\phi\rangle_B$ have infinite volume $V$, infinite degrees of freedom $N$ with a finite $N/V$.
\end{enumerate}%
%
%
Thermodynamic limit is needed because the holographic result \eqref{CproptoV1} is valid in that limit. Combining Eq.~\eqref{cforinner1} and \eqref{relCVABs2b}, we conclude that $f(x) \sim \ln x$ so
\begin{equation}\label{defholoCeq2}
  \bar{\C}(|\psi_1\rangle,|\psi_2\rangle)=- \ln|\langle\psi_1|\psi_2\rangle|^2\,,
\end{equation}
where the power $2$ is just our convention.
In fact, Eq.~\eqref{defholoCeq2} is a deformation of the Fubini-Study distance. For two pure states, the Fubini-Study distance is defined as
\begin{equation}\label{FBdist}
  D_{FS}(\psi_1\rangle,|\psi_2\rangle):=\arccos|\langle\psi_1|\psi_2\rangle|\,,
\end{equation}
so
\begin{equation}\label{defholoCeq200}
  \bar{\C}(|\psi_1\rangle,|\psi_2\rangle)=- 2\ln\cos[D_{FS}(\psi_1\rangle,|\psi_2\rangle)]\,.
\end{equation}
This deformation $D_{FS}\rightarrow\bar{\C}$ is a monotonically increasing function of Fubini-Study distance. In section \ref{answers1}, we will show  that above deformation can overcome the shortages of the Fubini-Study distance.
In the following subsections, we will demonstrate rich physics obtained from Eq. \eqref{defholoCeq2}.

\subsection{Path-integral formula and proof for path-integral complexity}
In this subsection, we will compute the complexity for pure states by the path integral formulation.
%
Suppose that $|\psi_0\rangle$ is a normalized initial state,  $|\psi(t)\rangle$ is a target state and the time evolution is give by a time evolution operator $\hat{U}(t)$.
Without loss of generality, we may consider the quantum mechanic case and assume that the configuration space is one dimensional.
The Feynman propagator reads,
\begin{equation}\label{pathforK1}
  K(x_2,t_2;x_1,0):=\langle x_2|\hat{U}(t)|x_1\rangle=\frac1{\mathcal{N}}\int_{x(0)=x_1}^{x(t_2)=x_2}\pD[x]\exp\left\{\frac{i}{\hbar}S[x(t)]\right\} \,,
\end{equation}
where $S[x(t)]$ is the classical action functional and $\mathcal{N}$ is the normalized factor.  The complexity between $|\psi_0\rangle$ and $|\psi(t)\rangle$ is given by Eq.~\eqref{defholoCeq2}:
\begin{equation}\label{waveCm1}
  \bar{\C}(t_2)=-\ln\left|\iint\td x_2\td x_2\psi_0^*(x_2)K(x_2,t_2;x_1,0)\psi_0(x_1)\right|^2\,,
\end{equation}
where $\psi_0(x):=\langle x|\psi_0\rangle$ is the wave function of the initial state. The time evolution of the complexity depends on the initial state and the action of the system.

The similar expression can be obtained in quantum field theory. The complexity between the state $|\Psi\rangle$ and $|\Phi\rangle$ in field theory can be expressed as a functional integration,
\begin{equation}\label{modCforscalar}
  \bar{\C}=-\ln\left|\int\pD[\varphi(x)]\Phi^*[\varphi(x)]\Psi[\varphi(x)]\right|^2\,,
\end{equation}
where $\Psi[\varphi(x)]=\langle\varphi|\Psi\rangle$ is the wave functional of the state $|\Psi\rangle$. For the time evolution case, the complexity between $|\Psi(t)\rangle$ and $|\Psi_0\rangle=|\Psi(0)\rangle$  can  be expressed as
\begin{equation}\label{CmPsi1}
\begin{split}
  \bar{\C}(t_2)&=-\ln\left|\int\pD[\varphi_1(x)]\pD[\varphi_2(x)]\Psi_0^*[\varphi_2(x)]\Psi_0[\varphi_1(x)]K[\varphi_2(x),t_2;\varphi_1(x),0]\right|^2\,,
  \end{split}
\end{equation}
where
\begin{equation}\label{pathintphi}
  K[\varphi_2(x),t_2;\varphi_1(x),t_1]=\frac1{\mathcal{N}}\int_{\varphi(x,t_1)=\varphi_1(x)}^{\varphi(x,t_2)=\varphi_2(x)}\pD[\varphi(x)]\exp\left\{\frac{i}{\hbar}S[\varphi]\right\}\,.
\end{equation}

In particular, we are interested in the complexity between the field operator eigenstate $|\varphi_0\rangle$ and the ground state $|\Omega\rangle$ for a given Hamiltonian.
The field operator eigenstate is the continuous limit of the product state in the configuration space, which is assumed as the reference state in path-integral complexity.
The inner product between these two states can be obtained by the Euclidean path integral:
\begin{equation}\label{innerOmega3}
\begin{split}
  e^{-\bar{\C}}=|\langle\varphi_0(x)|\Omega\rangle|^2=\frac1{\mathcal{N}}\int_{\varphi(x,0)=\varphi_0(x)}\pD[\varphi(x)]\exp\left\{-\frac{1}{\hbar}S_E[\varphi]\right\}\,,
  \end{split}
\end{equation}
where  $\varphi_0(x) = \langle x | \varphi_0  \rangle$, and $S_E[\varphi]$ is the Euclidean action, and {the normalization factor $\mathcal{N}$ is defined as
\begin{equation}
\mathcal{N}:=\int_{\varphi(x,0)=\Omega(x)}\pD[\varphi(x)]\exp\left\{-\frac{1}{\hbar}S_E[\varphi]\right\}\,.
\end{equation}
so that  $|\langle\Omega|\Omega\rangle|^2=1$.} The absolute symbol in the right-hand of Eq.~\eqref{innerOmega3} has been dropped as the function in the integration is positive definite. The upper bound of integration is omitted since, in the Euclidean case, the $\varphi(x,\infty)$ is the ground state $\Omega (x)= \langle x| \Omega \rangle$ and we do not need to specialize it.
In the classical limit $\hbar\rightarrow0$, the complexity between the ground state and a given eigenstate of the field operator is approximately
\begin{equation}\label{innerOmega3b}
  \bar{\C}\approx\frac{1}{\hbar}\min\{S_E[\varphi]-S_0 \   | \ \forall\varphi, \ s.t. \ \varphi(x,0)=\varphi_0(x)\}\,,
\end{equation}
where $S_0=\ln\mathcal{N}\approx S_E[\Omega]$ is the Euclidean on-shell action for the ground state ($|\phi_0\rangle = |\Omega \rangle$).

From Eq.~\eqref{innerOmega3}, we can prove the conjecture about the ``path-integral complexity'' proposed by Refs.~\cite{Caputa:2017urj,Caputa:2017yrh} as follows.
Let us consider a 2-dimensional conformal field theory embedded in a higher D-dimensional flat space ($2<D<25$), which contains arbitrary matter fields coupling with string worldsheet. The classical action reads,
\begin{equation}\label{CFTworldsheet}
  S:=(2\pi\alpha)^{-1}S_X+S_m[\varphi,g_{ab}]\,,
\end{equation}
where $S_X:=\int\td^2x g^{ab}\partial_aX^\mu\partial_bX^\nu\eta_{\mu\nu}$ and $S_m$ is the string worldsheet action and conformal matter fields action respectively. $\alpha$ is the string coupling constant. $\eta_{\mu\nu}$ is the Minkowski metric at the D-dimensional background space. $g_{ab}$ is the induced metric in the worldsheet.

The Euclidian action can be written as \cite{POLYAKOV1981207,DAS1989,Ginsparg:1993is}
\begin{equation}\label{CFTworldsheet}
  S_E =S_m[\varphi,\delta_{ab}]+\frac{1}{2\pi\alpha}(S_X[X^\mu,\delta_{ab}]+S_L[\phi,\delta_{ab}]+S_{gh}[b^{ab},c_a,\delta_{ab}])\,,
\end{equation}
where  $S_L[\phi,\delta_{ab}]:={c}/(24\pi)\iint\td^2 x\left[\eta^{ab}\partial_a\phi\partial_b\phi+\mu e^{2\phi}\right]$ is the Liouville action with the  central charge $c$ and $S_{gh}[b^{ab},c_a,\delta_{ab}]$ is the action for ghost fields.  Assume $|\varphi_0\rangle$ is one common eigenstate when $\varphi=\varphi_0$, $X^\mu=X^\mu_0$ and $g_{ab}^{(E)}=\delta_{ab}$; and $|\Omega_\phi\rangle$ is the ground state satisfying $g_{ab}^{(E)}|_{z=z_0}=e^{2\phi(x)}\delta_{ab}$, where $z$ is the Euclidean time and $z_0=\epsilon\ll1$ is a UV cut-off. Then we have
\begin{equation}\label{Liuo1s}
\begin{split}
  |\langle\varphi_0|\Omega_\phi\rangle|^2=&\int\pD[\phi]\pD[\varphi]\pD[X]\pD[b]\pD[c]\exp\left\{-\frac{1}{\hbar}S_E\right\}\\
  =&\left[\int\pD[\phi]\exp\left(-\frac{S_L}{2\pi\alpha\hbar}\right)\right]|\langle\varphi_0|\Omega_0\rangle|^2\,,
  \end{split}
\end{equation}
where $|\Omega_0\rangle$ is the ground state when $\phi=0$. Thus, the complexity between $|\varphi_0\rangle$ and $|\Omega_\phi\rangle$ reads
\begin{equation}\label{Cmforphi1}
  \bar{\C}[\phi]=-\ln\int_{\phi(x,z=\epsilon)=\phi(x)}\pD[\phi]\exp\left(-\frac{S_L}{2\pi\alpha\hbar}\right) -\ln|\langle\varphi_0|\Omega_0\rangle|^2\,,
\end{equation}

It is interesting to compare our result Eq.~\eqref{Cmforphi1} with the proposal of the path integral complexity in Refs.~\cite{Caputa:2017urj,Caputa:2017yrh}. Refs.~\cite{Caputa:2017urj,Caputa:2017yrh} conjectured that the complexity between ground state $|\Omega\rangle$ and the field operator eigenstate $|\varphi_0\rangle$ was given by the on-shell Liouville action.

{In the small $\hbar \alpha$ limit\footnote{There are two different limits that we can recover the proposal about Liouville action: $\hbar\rightarrow0$ and $\alpha\rightarrow0$. The former  corresponds to the usual classical limit while the later corresponds to the weak coupling limit between the matter and string/gravity.}, the saddle point approximation of Eq.~\eqref{Cmforphi1} yields
\begin{equation}\label{Cmforliovs2}
  \bar{\C}=\bar{\C}(0)+\frac{S_L^{(cl)}[\phi]}{2\pi\hbar\alpha}[1+\mathcal{O}(\hbar\alpha)]\,,
\end{equation}
}
where $S_L^{(cl)}[\phi]$ is the classical on-shell action of the Liouville action with the boundary condition $\phi(x,\epsilon)=\phi(x)$.  Thus, we see that the conjecture of Refs.~\cite{Caputa:2017urj,Caputa:2017yrh} only includes the leading order term in classical limit. $\bar{\C}(0)$ corresponds to $S_0$ in Eq \eqref{innerOmega3b}.

Ref.~\cite{Czech:2017ryf} also gave a more exact diagrammatic argument about why $\bar{\C}$ should be proportional to $S_L$ by the relationship between discretized path integrals and tensor network renormalization algorithm~\cite{PhysRevLett.115.180405}.
Our result is purely algebraic and the starting point has no relationship with the tensor network renormalization. This agreement, notwithstanding the different method, is an evidence supporting our proposal.

\subsection{Relation to holographic conjectures}
In this section, we will show that both the CV and CA conjectures can be understood from our proposal, which serves as another supporting evidence for our proposal.  We also clarify what the reference states are in these two conjectures and why two conjectures have different behaviors at early time~\cite{Carmi:2017jqz,Kim:2017qrq}.

\subsubsection{CV conjecture}
Let us consider the ground state $|\Omega\rangle$ for a given CFT Hamiltonian $H_0$. Let us  make a perturbation on this Hamiltonian $H'=H_0+ H_I\delta$ with an infinitesimal parameter $\delta$ and obtain the perturbed ground state $|\Omega_\delta\rangle$. Then the fidelity (Fi) between these two ground states reads
\begin{equation}\label{fi2Omega}
  \Fi(|\Omega\rangle,|\Omega_\delta\rangle)=|\langle\Omega|\Omega_\delta\rangle|=1-G_{\delta\delta}\delta^2+\mathcal{O}(\delta^4) \,,
\end{equation}
where $G_{\delta\delta}$ is called the information metric~\cite{MIyaji:2015mia} or fidelity of susceptibility~\cite{doi:10.1142/S0217979210056335}. Neglecting the higher order of $\delta$ we find that the complexity between  $|\Omega\rangle$ and  $|\Omega_\delta\rangle$ is
\begin{equation}\label{CforOmega}
  \bar{\C}(|\Omega\rangle,|\Omega_\delta\rangle)=-2\ln\Fi(|\Omega\rangle,|\Omega_\delta\rangle)=2G_{\delta\delta}\delta^2 \,.
\end{equation}
Thus, we find a simple relationship between the complexity of the perturbed ground states and the information metric
\begin{equation}\label{relC2IM}
  \bar{\C}(|\Omega\rangle,|\Omega_\delta\rangle)\propto G_{\delta\delta}\,.
\end{equation}

Refs.~\cite{MIyaji:2015mia,Alishahiha:2017cuk} have given some nontrivial evidence to show that, in conformal field theories with a small perturbation by a primary operator, the information metric is approximately given by a volume of the maximal time slice in the AdS spacetime, i.e.,
\begin{equation}\label{IMVs1}
  G_{\delta\delta}\propto\max_{\partial \Sigma=t_L\cup t_R}\text{Vol}(\Sigma)\,.
\end{equation}
Thus, given that the holographic duality~\eqref{IMVs1} is correct, we obtain by Eq. \eqref{relC2IM},
\begin{equation}\label{IMVs2}
 \bar{\C}(|\Omega\rangle,|\Omega_\delta\rangle)\propto\max_{\partial \Sigma=t_L\cup t_R}\text{Vol}(\Sigma) \,.
\end{equation}
This is nothing but the CV conjecture! The ground state of a CFT in holography is the TFD state dual to the double-sided black hole geometry. Thus, the complexity in the CV conjecture is the complexity between the TFD state and its perturbed TFD state under a marginal operator, not the complexity between a TFD state and an unknown ``simple'' reference state, which is usually assumed in most literatures.

\subsubsection{CA conjecture}
Let us turn to the CA conjecture.
Firstly, from Eq.~\eqref{innerOmega3} we see that the complexity between the  field operator eigenstate and the ground state of a given Hamiltonian is given by the partition function of the boundary field theory
\begin{equation}\label{C2Zbd1}
  \bar{\C}=-\ln Z_{\text{bd}}[\phi(x)] \,.
\end{equation}
On the other hand, the partition function of the boundary field theory in AdS/CFT correspondence is given by the partition function of a bulk gravity theory in asymptotic AdS spacetime
\begin{equation}\label{AdSCFTZ1}
  Z_{\text{bd}}[\phi(x)]=Z_{\text{bulk}}[g_{\mu\nu},\phi(x,z)]\,,
\end{equation}
with matter fields which satisfy the boundary condition $\phi(x,z)|_{z=0}=\phi(x)$. 
Then we can find that Eq.~\eqref{C2Zbd1} reads,
\begin{equation}\label{C2Zbd2}
  \bar{\C}=-\ln \int\pD[g_{\mu\nu}]\pD[\phi]\exp\left\{-\frac1{\hbar}S_E[g_{\mu\nu}\,,\phi(x,z)]\right\}\,.
\end{equation}
where $S_E$ is the Euclidian action of the bulk gravity with matters.

In the weak gravity  limit, we have the following leading order approximation
\begin{equation}\label{C2Zbd3}
  \bar{\C}\approx\frac1{\hbar}S_{E,\text{on-shell}}[g_{\mu\nu},\phi(x,z)]\,.
\end{equation}
In the Lorentz signature we have, by the Wick's rotation,
\begin{equation}\label{C2Zbd4}
  \bar{\C}\approx\frac1{\hbar}S_{\text{on-shell}}[g_{\mu\nu},\phi(x,z)]=\frac1{\hbar}\left[\int_{-\infty}^\infty\td t\int_{V(t)}\mathcal{L}(g_{\mu\nu},\phi)\td^dx+S_{\text{bd}}\right]\,,
\end{equation}
where $V(t)$ is a time slice in the bulk at time $t$, $\mathcal{L}(g_{\mu\nu},\phi)$ is the Lagrangian density of the gravity theory with bulk matters and $S_{\text{bd}}$ is a suitable boundary term. We assume that the bulk spacetime is $(d+1)$-dimensional. In order to compute the integration~\eqref{C2Zbd4}, we need to carefully define the integration region.  The time slices $V(t)$ should satisfy a suitable boundary condition so that the bulk region $\cup_{t\in\mathbb{R}}V(t)$ can correspond to the boundary states given by $t_L$ and $t_R$.
Thus, we require the time slices $V(t)$ satisfies the following boundary condition
\begin{equation}\label{boundaryCD}
  \forall t, ~~\partial V(t)=t_L\cup t_R\,.
\end{equation}
\begin{figure}
  \centering
  \includegraphics[width=0.8\textwidth]{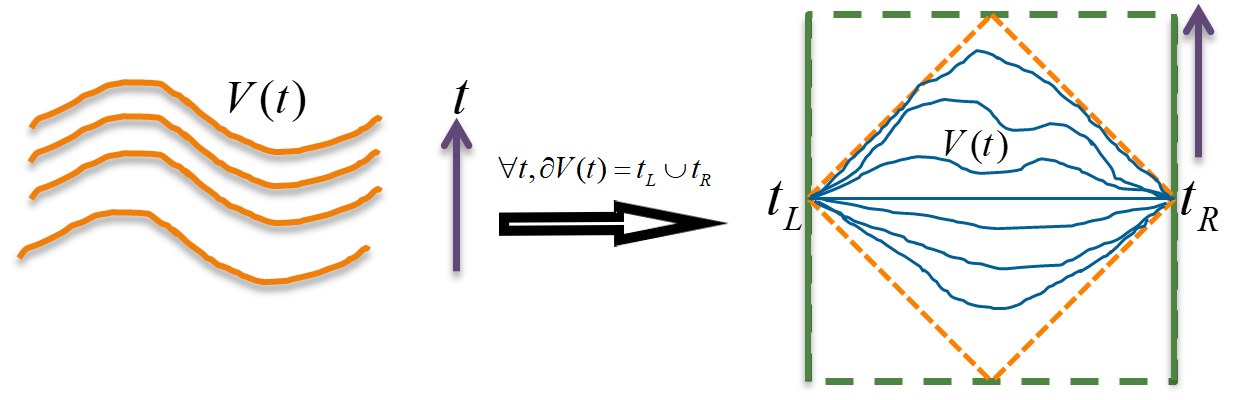}
  \caption{After we fix the boundary condition for the time slice $\partial V(t)=t_L\cup t_R$, the bulk region to compute the on-shell action is just the WdW patch.  } \label{CAbulk}
\end{figure}
From Fig.~\ref{CAbulk} we  find
\begin{equation}\label{C2Zbd5}
  \bar{\C}\approx\frac1{\hbar}\left[\int_{\text{WdW}}\mathcal{L}(g_{\mu\nu},\phi)\td^{d+1}x+S_{\text{bd}}\right]\,.
\end{equation}
This is nothing but the CA conjecture!

We see that the complexity in the CA conjecture describes the complexity between a ground state and the field operator eigenstate of a boundary field theory.
In the holographic duality, the ground state of the boundary field theory is the TFD state corresponding to the double-sided AdS black hole. Thus, the CA conjecture gives the complexity between a TFD state and the eigenstate of the field operator of the boundary field theory.
Note that it is not the complexity between two TFD states. This explains why the CV and CA conjecture show very different time-evolution behaviors at early time, which was reported in~\cite{Carmi:2017jqz,Kim:2017qrq}.

\section{Applications to chaotic systems}\label{app-chaos}

In this section we apply our proposal \eqref{defholoCeq2}  to chaotic systems.  We find an interesting relation between the Lyapunov exponent and the complexity growth rate at late time.

\subsection{Relation between the Lyapunov exponent and the complexity}
In this subsection, we will present an interesting relation between the Lyapunov exponent  and the time evolution property of the complexity.

In order to give an intuition how it works, let us first consider an ``imaginary frequency'' harmonic oscillator. The Lagrangian for such a system is given by
\begin{equation}\label{oscillator1}
  \mathcal{L}=\frac{m^2}2\dot{x}^2+\frac{k^2}2x^2\,,
\end{equation}
where $\omega$ and $m$ are positive. This Lagrangian gives the following Hamiltonian
\begin{equation}\label{oscillator2}
  H=\frac{p^2}{2m}-\frac{k^2}2x^2\,.
\end{equation}
Note that this Hamiltonian is not bounded below so the quantum mechanics is ill-defined. Although the classical mechanics of this system gives a positive Lyapunov exponent, the system is not chaotic since it is a linear dynamical system. However, it is a good ``toy'' model to exhibit the connection between the complexity growth rate and the Lyapunov exponent. After having some intuition form this toy model, we will move on to well-defined chaotic quantum systems.

Classical chaotic systems are characterized by the Lyapunov exponents. For a general 1-dimensional dynamic system, its motion in phase space is given by $(x(t),p(t))$ with initial point $(x_0,p_0)$. Suppose that, at the initial time, the starting point at the phase space has an infinitesimal deviation. This will generat a new trajectory $(\tilde{x}(t),\tilde{p}(t))$. The deviation between these two trajectories can be characterized by a function
\begin{equation}
\gamma(t):=\sqrt{[x(t)-\tilde{x}(t)]^2+[p(t)-\tilde{p}(t)]^2} \,.
\end{equation}
The Lyapunov exponent for the initial point $(x_0,p_0)$ is defined as
\begin{equation}\label{LEx0p0}
  \lambda_L(x_0,p_0):=\lim_{t\rightarrow\infty}\lim_{\gamma(0)\rightarrow0}\frac1t\ln\frac{\gamma(t)}{\gamma(0)} \,.
\end{equation}
The equation of motion from the Lagrangian~\eqref{oscillator1} reads
\begin{equation}\label{eqforosci1}
  \ddot{x}-\omega^2x=0\,,
\end{equation}
where $\omega := \sqrt{k/m}$.
The Lyapunov exponents \eqref{LEx0p0} for this system are
\begin{equation} \label{Lyap11}
\lambda_L=\omega \,,
\end{equation}
and $\lambda_L^{(-)}=-\lambda_L$.

Now let us  compute the complexity between an initial state $|\psi_0\rangle$ and its time evolution state $|\psi(t)\rangle$. The Feynman propagator can be computed analytically,
\begin{equation}\label{pathforKs1}
  K(x_2,t;x_1,0)=\sqrt{\frac{m\omega}{2\pi i\hbar\sinh\omega t}}\exp\left[\frac{i}{\hbar}S_{\text{cl}}(x_2,t,x_1)\right] \,,
\end{equation}
where $S_{\text{cl}}(x_2,t,x_1)$ is the on-shell action for the trajectory $x(t)$ which satisfies the conditions $x(0)=x_1$ and $x(t)=x_2$:
\begin{equation}\label{eqforS1}
  S_{\text{cl}}(x_2,t,x_1)=\frac{m\omega}{2\sinh\omega t}[(x_2^2+x_1^2)\cosh\omega t-2x_1x_2] \,.
\end{equation}
Plugging this into Eq.~\eqref{waveCm1} we obtain the complexity between an initial state and its time evolution state in late time limit:
%
\begin{equation}\label{waveCm1s12}
  \bar{\C}(t)=\lambda_Lt+\text{finite constant} \,,
\end{equation}
where we used $\sinh\omega t\rightarrow e^{\omega t}/2=e^{\lambda_Lt}/2$ with Eq. \eqref{Lyap11}.
We find that the complexity grows linearly and the slope (growth rate) is just the positive Lyapunov exponent.\footnote{Strictly speaking, as we have a freedom in choosing the overall factor in Eq.~\eqref{defholoCeq2}, the growth rate of the complexity will be proportional to the Lyapunov exponent.}
Though the definition of the complexity includes an initial state $|\psi_0\rangle$, the complexity at late time is independent of the initial state $|\psi_0\rangle$ and only has something to do with the Lyapunov exponent.

Let us now move on to a more general case with the Hamiltonian
\begin{equation}\label{Lform1}
  H(x,p;t)=\frac{p^2}{2m}+V(x,t) \,,
\end{equation}
where $p$ is the canonical momentum, $m$ is the mass and $V(x,t)$ is an arbitrary bounded below potential. In addition $V(x,t)$ has a higher order term than a quadratic term so that we can have a chaotic behavior. We may study the time evolution of the complexity for this system in the semiclassical situations. The analysis is straightforward but tedious so we relegate details to subsection~\ref{chaoticC1}.
In summary, we find very interesting results:
\begin{enumerate}
\item[(1)] The complexity grow linearly at late time if the classical system is chaotic. The growth rate is determined by the {\it Lyapunov exponent}\footnote{We refer to~\cite{Jahnke:2018off, Jahnke:2019gxr} and references therein for more extensive holographic discussions of the quantum chaos and the Lyapunov exponent.} (Eq.~\eqref{complexL1})
\begin{equation}\label{complexL11}
  \bar{\C}(t)={\lambda}_Lt+\cdots\,.
\end{equation}
We emphasize that this is a {\it proof} of an important {\it proposed} property of the complexity in many literatures. See subsection~\ref{chaoticC1} for more details.
\item[(2)]
This linear-in-time behavior breaks down at the time scale $t \sim t_{cl}$ (Eq.~\eqref{logtime})
\begin{equation}\label{logtime11}
t_{cl}:=-\frac{1}{2{\lambda}_L}\ln(c_1\hbar) \,,
\end{equation}
where $c_1$ is model-dependent constant. The time scale $t_{cl}$ is the \textit{log time barrier} or usually called ``Ehrenfest time''~\cite{Ehrenfest1927,PhysRevE.65.035208,Schubert_2012}.
\item[(3)]  For field theory systems Eq.\eqref{complexL11} can be understood as (Eq.~\eqref{complexL2})
\begin{equation}\label{complexL211}
  \dot{\bar{\C}}(t)\propto E+\cdots\,,
\end{equation}
which is similar to the growth rate of the complexity in the CV and CA conjectures.
\end{enumerate}

Note that Eqs.~\eqref{complexL11} and \eqref{complexL211} were conjectured in many literatures but  were not proven explicitly. Here, we demonstrate how Eq.~\eqref{complexL11}  arises by using our proposal \eqref{defholoCeq2}.  In addition, we are also able to show when such a linear growth should terminate.
Furtheromre, in \cite{Yang:2019iav} we explicitly demonstrate this predicted time evolution in concrete examples. They support our claim that the complexity may be unitary-invariant.

\subsection{Linear growth of the complexity in chaotic systems: a proof}\label{chaoticC1}
In this subsection, we explain Eq.~\eqref{complexL11} - \eqref{complexL211} in more detail and show a very important result: the complexity grows linearly at late time if the classical system is chaotic. This property was proposed in many discussions from the holographic duality but it was not proven yet in general. In addition to this property, we also show that such linear growth should end at the critical time so called a log time barrier.

Let us consider the Hamiltonian
\begin{equation}\label{Lform1}
  H(x,p;t)=\frac{p^2}{2m}+V(x,t) \,,
\end{equation}
where $p$ is the canonical momentum, $m$ is the mass and $V(x,t)$ is an arbitrary bounded below potential. We study the time evolution of this system in the semiclassical situations.

When the quantum fluctuation is not very strong, we can use the Van Vleck-Pauli-Morette formula to compute the semiclassical result of the Feynman propagator, which is
%
\begin{equation}\label{pathforK2}
\begin{split}
  K(x_2,t_2;x_1,0)\approx\sqrt{\frac{m}{2\pi i\hbar J(x_2,t_2,x_1)}}\exp\left\{\frac{i}{\hbar}S[x_{cl}(t)]\right\}[1+\mathcal{O}(\hbar)]\,,
  \end{split}
\end{equation}
where $x_{cl}(t)$ is the classical trajectory with $x_{cl}(0)=x_1$ and $x_{cl}(t_2)=x_2$ ,  and $J(x_2,t_2,x_1)$ is the Jacobi field which satisfies equation
\begin{equation}\label{Jaccobieq1}
  m^2\frac{\td }{\td t^2}J(x_2,t,x_1)+J(x_2,t_2,x_1)\partial_x^2V(x_{cl} \,,(t),t)=0\,.
\end{equation}
with the initial condition
\begin{equation}
J(x_2,0,x_1)=0, \left.\frac{\td}{\td t}J(x_2,t,x_1)\right|_{t=0}=1\,.
\end{equation}
Note that if the action is quadratic form of momentum $\hat{p}$ and position $\x$, the expression~\eqref{pathforK2} is the exact result and there is no correction term $\mathcal{O}(\hbar)$. However, such an action is not chaotic because it is a linear system.

As the classical trajectory is the function of $x_1$ and $x_2$, the Jacobi field and the on-shell action $S[x_{cl}]$
\begin{equation}
S[x_{cl}]=S_{cl}(x_2,t_2,x_1)\,,
\end{equation}
are functions of $x_1$ and $x_2$.
It is assumed that there is no conjugate point for $t\in(0,t_2)$, i.e., $J(x_2,t,x_1)\neq0$ for $0<t<t_2$.


Since we consider the  system in the semiclassical approximation, it is natural to assume that the initial wave function is the following wave package
\begin{equation}\label{Gaussianwave}
  \psi_0(x):=\langle x|\psi_0\rangle\approx\frac1{\sqrt{\sqrt{\hbar}\sigma}}f[x/(\sigma\sqrt{\hbar})]\,,
\end{equation}
where $f$ is any function which rapidly decays  when $(x-x_0)/(\sigma\sqrt{\hbar})\rightarrow\pm\infty$ and also $|f|^2\rightarrow\sqrt{\hbar}\sigma\delta(x-x_0)$   when $\hbar\rightarrow0$.
%
Here, $\sigma$ is the width of the wave package and $x_0$ is the position of the particles at classical limit. 

For a chaotic system, when $t_2$ is large, the solution of the Jacobi field will grow exponentially and always have the form
\begin{equation}\label{resultofJ}
  J(x_2,t_2,x_1)=J_0(x_2,t_2,x_1)e^{t_2\lambda_L(x_2,x_1)}\,,
\end{equation}
with the Lyapunov exponent $\lambda_L(x_2,x_1)>0$ for given $\{x_1, x_2\}$.  A smooth function $J_0(x_2,t_2,x_1)$ is bounded or a polynomial function of $t_2$ for large $t_2$.
Thus, we see that the complexity between $|\psi(t)\rangle$ and $|\psi_0\rangle$ \eqref{waveCm1} reads
\begin{equation}\label{waveCm4b}
\begin{split}
  e^{-\bar{\C}(t_2)}&\propto\left|\iint\td x_2\td x_1h(x_2,t_2,x_1)e^{-\lambda_L(x_2,x_1)t_2/2}\exp\left[\frac{i}{\hbar}S_{\text{cl}}(x_2,t_2,x_1)\right]\right|^2\,,
  \end{split}
\end{equation}
with
\begin{equation}\label{definechi}
\begin{split}
  h(x_2,t_2,x_1)=f[x_1/(\sigma\sqrt{\hbar})]f^*[x_2/(\sigma\sqrt{\hbar})]J_0(x_2,t_2,x_1)^{-\frac12}\,,
  \end{split}
\end{equation}
which is a smooth, bounded and rapidly decaying function for all $t_2$.

As the function $\exp\left[\frac{i}{\hbar}S_{\text{cl}}(x_2,t_2,x_1)\right]$
is rapidly oscillating function when $\hbar\rightarrow0$ we can use the stationary phase approximation to find the leading term of   Eq.~\eqref{waveCm4b}.
The stationary phase points $\{x_1=a_k,x_2=b_k\}$ (with $k=1,2,\cdots$) are given by following equations,
\begin{equation}\label{expandvarps}
   \left.\frac{\partial}{\partial x_1}S_{\text{cl}}(x_2,t_2,x_1)\right|_{x_1=a_k,x_2=b_k}=\left.\frac{\partial}{\partial x_2}S_{\text{cl}}(x_2,t_2,x_1)\right|_{x_1=a_k,x_2=b_k}=0 \,,
\end{equation}
and the restriction to the determinants
\begin{equation}
A_{k}:=\det\left[\frac{\partial^2S_{\text{cl}}(x_2,t_2,x_1)}{\partial x_i\partial x_j}\right]_{x_1=a_k,x_2=b_k}\neq0,~~~i,j=1,2\,.
\end{equation}
The stationary phase approximation shows that if $\hbar\rightarrow0$
\begin{equation}\label{waveCm4c}
\begin{split}
  &\iint h(x_2,t_2,x_1)e^{-\frac{\lambda_L(x_2,x_1)t_2}2}\exp\left[\frac{i}{\hbar}S_{\text{cl}}(x_2,t_2,x_1)\right]\td x_2\td x_1\\
  &\propto\sum_{k}h(b_k,t_2,a_k)e^{-\frac{\lambda_L(b_k,a_k)t_2}2}\exp\left[\frac{i}{\hbar}S_{\text{cl}}(b_k,t_2,a_k)\right]\frac{e^{\frac{\pi}4 \text{sign}(A_k)i}}{\sqrt{|A_k|}}\,,
  \end{split}
\end{equation}

Suppose that $k=k_0$ can make $\lambda_L(a_k,b_k)$ minimal in all the saddle points and define
\begin{equation}
\bar{\lambda}_L:=\min\{\lambda_L(a_k,b_k)~|~k=1,2,\cdots\}=\lambda_L(b_{k_0},a_{k_0})>0\,.
\end{equation}
At the large time limit, the $k=k_0$ term will dominate the sum in Eq.~\eqref{waveCm4c}  so we have
\begin{equation}\label{waveCm4d}
\begin{split}
  &\iint\td x_2\td x_1h(x_2,t_2,x_1)e^{-\frac{\lambda_L(b_k,a_k)t_2}2}\exp\left[\frac{i}{\hbar}S_{\text{cl}}(x_2,t_2,x_1)\right]\\
  \propto& h(b_{k_0},t_2,a_{k_0})e^{-\frac{\bar{\lambda}_Lt_2}2}\exp\left[\frac{i}{\hbar}S_{\text{cl}}(b_{k_0},t_2,a_{k_0})\right]\frac{e^{\frac{\pi i}4 \text{sign}(A_{k_0})}}{\sqrt{|A_{k_0}|}}\,,
  \end{split}
\end{equation}
This implies that the complexity have the following asymptotic behavior at  large time $t$
\begin{equation}\label{complexL1}
  \bar{\C}(t)=\bar{\lambda}_Lt+\cdots\,.
\end{equation}
Thus, we have showed that the complexity between $|\psi(t)\rangle$ and $|\psi_0\rangle$ will grow linearly at the late time limit for $\hbar \ll 1$ and the system is chaotic.

One should keep in mind that the above results are obtained by the assumption that $\mathcal{O}(\hbar)$ in Eq.~\eqref{pathforK2} can be neglected. When we take the first order quantum effects into account,  the above results will not be valid if
\begin{equation}\label{logtime}
  t\gtrsim t_{cl}:=-\frac{1}{2\bar{\lambda}_L}\ln(c_1\hbar) \,,
\end{equation}
where $c_1$ is a model-dependent constant. The time scale $t_{cl}$ is the \textit{log time barrier} or usually called ``Ehrenfest time''~\cite{Ehrenfest1927,PhysRevE.65.035208,Schubert_2012}, below which the complexity growth for chaotic systems can be obtained by its classical chaotic behaviors.  When $t>t_{cl}$, we have to solve the Feynman propagator beyond the classical order and quantum effects matters. In general, the complexity between $|\psi(t)\rangle$ and $|\psi_0\rangle$  will not grow linearly if $t>t_{cl}$.

We may generalize the above results to the $d$-dimensional space with $N$ particles, which can be regarded as a one-particle system in the $dN$-dimensional space.
Suppose that the position of this system is given by $\vec{x}:=(x^{(1)},x^{(2)},\cdots,x^{(dN)})$.  The propagator in the semiclassical limit reads
\begin{equation}\label{pathforK2b}
\begin{split}
  K(\vec{x}_2,t_2;\vec{x}_1,0)\approx\sqrt{\frac{m}{2\pi i\hbar \det[J_{kl}(\vec{x}_2,t_2,\vec{x}_1)]}}\exp\left\{\frac{i}{\hbar}S[\vec{x}_{cl}(t)]\right\}\,,
  \end{split}
\end{equation}
and the Jacobi field becomes a $dN$-dimensional matrix satisfying the  equation
\begin{equation}\label{Jaccobieq1}
  m^2\frac{\td }{\td t^2}J_{kl}(\vec{x}_2,t_2,\vec{x}_1)+[\partial_{k}\partial_{j}V(x_{cl}(t),t)]J_{jl}(\vec{x}_2,t_2,\vec{x}_1)=0\,,
\end{equation}
with the initial conditions
\begin{equation}
J_{kl}(\vec{x}_2,0,\vec{x}_1)=0, \left.\frac{\td}{\td t}J_{kl}(\vec{x}_2,t_2,\vec{x}_1)\right|_{t=0}=\delta_{kl}\,,
\end{equation}
where $\partial_k:=\partial/\partial x^{(k)}$.

If the effective Lyapunov exponent of a single particle in 1-dimensional case is $\bar{\lambda}_L$, then
\begin{equation}
\det[J_{kl}(\vec{x}_2,t_2,\vec{x}_1)]\propto\exp(dN\bar{\lambda}_L t)\,.
\end{equation}
Thus, the complexity growth rate at late time limit is
\begin{equation}\label{complexL2}
  \bar{\C}(t)=dN\bar{\lambda}_Lt+\cdots\,.
\end{equation}
As the total energy $E$ is proportional to the particles numbers when $N$ is large, we have
\begin{equation}\label{complexL2}
  \dot{\bar{\C}}(t)\propto E+\cdots\,,
\end{equation}
which is very similar to the growth rate of the complexity in the CV and CA conjectures.

\section{Comments on unitary-invariant complexity}\label{answers1}

In this section, we make two comments on our proposal. First, we discuss how our proposal can resolve the problem in the Fubini-Study distance. Second, we apply our proposal to the TFD states and show our proposal is consistent with the holographic results.
\subsection{Difference from the Fubini-Study distance}

In Ref.~\cite{Brown:2017jil} it was argued that that the Fubini-Study distance~\eqref{FBdist} could not be the complexity because it cannot distinguish one-flip from multi-flips.  However, this should not be considered as an objection to using the inner product. Our purpose here is to show that this problem (distinguishing one-flip from multi-flips) may be solved by our proposal with the inner product\eqref{defholoCeq2}. i.e. by using ``$\ln$'' instead ``arccos''.

We start with the problem of \eqref{FBdist} as the complexity.  For example, let us consider the following two states in a $n$-qubit system
\begin{equation}\label{qubitstate1}
  |\psi_1\rangle=|a_1\rangle\otimes|a_2\rangle\otimes\cdots\otimes|a_n\rangle \,,
\end{equation}
and
\begin{equation}\label{qubitstate2}
|\psi_2\rangle=|b_1\rangle\otimes|b_2\rangle\otimes\cdots\otimes|b_n\rangle \,,
\end{equation}
with $a_i=0,1$ and $b_i=0,1$. Let us first consider the Fubini-Study distance \eqref{FBdist}:
\begin{equation}\label{FBdist2}
  D_{FS}=\arccos\left|\prod_{i=1}^n\langle a_i|b_i\rangle\right|\,.
\end{equation}
%
{If we start with $|\psi_1\rangle = |\psi_2 \rangle$} we can flip  only one qubit to change the complexity from {zero} into $\pi/2$. This does not reflect the property of the complexity as we expect changing just one qubit should not change the complexity too much if the system is large enough. {If we start with $\langle \psi_1 | \psi_2 \rangle = 0$, flipping some of the qubit will not change the complexity at all. } This again does not reflect the property of the complexity: the state is changing but the complexity is not.

Next, let us consider our proposal:
\begin{equation}\label{nqubitsum}
  \bar{\C}=-2\ln|\langle\psi_1|\psi_2\rangle|=-2\sum_{i=1}^n\ln|\langle a_i| b_i\rangle|\,.
\end{equation}
Here, we do not have the problem in the Fubini-Study distance.  Thanks to ``log'', naively every flip will have one-unit of the complexity, which is a desired property for the complexity. However, in this case, one-unit of the complexity seems infinite. This infinity problem can be resolved if we note that our formula~\eqref{defholoCeq2} are proposed for the system in the \textit{continuous and thermodynamics limit} rather than finite discrete systems.


For a continuous 1 dimensional system
\begin{equation}\label{contqubit11}
\bar{\C}=-2\ln|\langle\psi_1|\psi_2\rangle|=-2L\int\ln|\langle a(k)|b(k)\rangle|\td k\,,
\end{equation}
where the discrete index $i$ is replaced by the continuous label $k$ and $L$ has the dimension of $[k]^{-1}$.  {Let us assuem the states are `regular' states, which means that the inner product
\begin{equation}\label{regularassum1}
  \langle a(k)|b(k)\rangle~~\text{is the analytical function of}~k\,.
\end{equation}
If we change the states of $k\in(k_0-\delta,k_0+\delta)$ so that $\langle a(k_0)|b(k_0)\rangle\neq0\rightarrow\langle \tilde{a}(k_0)|\tilde{b}(k_0)\rangle=0$, the change of the complexity is
\begin{equation}\label{contqubit1b}
\delta\bar{\C}=-2L\int_{-\delta}^{\delta}[\ln|\langle \tilde{a}(k_0+x)|\tilde{b}(k_0+x)\rangle|-\ln|\langle a(k_0+x)|b(k_0+x)\rangle|]\td x\,.
\end{equation}
This integration is finite again thanks to `log', although  $\langle \tilde{a}(k_0+x)|\tilde{b}(k_0+x)\rangle$ is zero at $x=0$. In the limit of $\delta\rightarrow0$, which correspondss to ``flipping exactly one qubit'',  $\delta\bar{\C}=0$ as expected. }
%

If we try to use our proposal for the discrete system, such as the $n$-qubit system, we have to make a suitable regularization in the argument of ``$\ln$'' and a suitable discretization on the integration measure ``$\int\td k$''. For a $n$-qubit system, a convenient method is that
\begin{equation}
L\int\td k\rightarrow\sum \,,
\end{equation}
for a discretization and
\begin{equation} \label{coff123}
\ln|\langle\cdot|\cdot\rangle|\rightarrow\ln(|\langle\cdot|\cdot\rangle|+\bar{\varepsilon}) \,,
\end{equation}
with $\epsilon\ll1$ for a regularization. Then for the two states Eq.~\eqref{qubitstate1} and \eqref{qubitstate2} in the $n$-qubit system, the discrete version of Eq.~\eqref{contqubit11} is
\begin{equation}\label{contqubit111}
  \bar{\C}(|\psi_2\rangle,|\psi_1\rangle)=-2\sum_{i=1}^n\ln\left(\frac{|\langle a_i|b_i\rangle|+\bar{\varepsilon}}{1+\bar{\varepsilon}}\right) \,,
\end{equation}
where the denominator $1+\bar{\varepsilon}$ is introduced to ensure that the complexity between the same states is zero. Suppose that we start with the case $|\psi_1\rangle = |\psi_2 \rangle$, so $\bar{\C}(|\psi_2\rangle,|\psi_1\rangle) =0$. If we flip only one qubit in $|\psi_2 \rangle$, then the complexity will change by $\bar{\C}_0=-2\ln(\bar{\varepsilon}/(1+\bar{\varepsilon})) \sim -2\ln \bar{\varepsilon} $. If we flip $n$ qubits in $|\psi_2 \rangle$ the complexity is changed by $n\bar{\C}_0$, which is a desired property of the complexity.

At first glance, the cut-off term $\bar{\varepsilon}$ in Eq.~\eqref{coff123} looks artificial. However, this may be understood as follows. Suppose that we want to create two qubits $|a_{\text{th}}\rangle$ and $|b_{\text{th}}\rangle$. 
Mathematically, it is easy to write down but physically we have to use some physical systems to realize them. Due to unavoidable quantum and thermal fluctuations, what we really observe are two states $|a_{\text{ob}}\rangle$ and $|b_{\text{ob}}\rangle$ and their inner product is
\begin{equation}
\overline{|\langle a_{\text{ob}}|b_{\text{ob}}\rangle|}=|\langle a_{\text{th}}|b_{\text{th}}\rangle|+\bar{\varepsilon}\,,
\end{equation}
where the ``$\overline{X}$'' stands for the average observations of the variable $X$ and $\bar{\varepsilon}$ is an ``error'' which is due to the intrinsic quantum and thermal fluctuations. This is one interpretation of  Eq.~\eqref{coff123}.

\subsection{Application to the TFD states and compatibility with holographic results}

We also would like to make a short comment regarding the inner product between a TFD state and its time-evolution state at large time limit:
\begin{equation}\label{defFts100}
  F(t):=|\langle\text{TFD}(0)|\text{TFD}(t)\rangle|\,,
\end{equation}
where a time-dependent TFD state
\begin{equation}\label{timeTFDstate00}
  |\text{TFD}(t)\rangle:=\frac1{\sqrt{Z(\beta)}}\sum_{n}e^{-(\beta+2it)E_n/2}|E_n\rangle_R|E_n\rangle_L\,,
\end{equation}
with the inverse temperature $\beta$, eigen-energy $E_n$ and partition function $Z(\beta)$. Ref.~\cite{Hashimoto:2018bmb} stated that, at late time limit, $F(t)$ would approach to a finite constant
\begin{equation}\label{wronglartget}
  \lim_{t\rightarrow\infty}F(t)\sim \mathrm{constant} \neq0\,,
\end{equation}
This seems to show that the $-\ln F(t)$ will not growth linearly when time is large and impossible to reproduce the expected behavior of complexity. In following, we will show this is not true.

Plugging Eq.~\eqref{timeTFDstate00} into Eq.~\eqref{defFts100} we obtain
\begin{equation}\label{innerTFDs1}
  F(t)=\frac1{Z(\beta)}\left|\sum_{n}e^{-\beta E_n}e^{iE_nt}\right|=\frac1{Z(\beta)}\left|Z(\beta+it)\right| \,,
\end{equation}
with
\begin{equation}
Z(\beta+it):=\sum_{n}e^{-\beta E_n}e^{iE_nt}\,.
\end{equation}
In the continuous limit, we may replace the sum with the integral:
\begin{equation}\label{intforfTFD}
  Z(\beta+it)=\int_0^\infty N(E)e^{-\beta E}e^{iEt}\td E\,,
\end{equation}
where the density of state $N(E)$ is introduced and $N(E)\td E$ is the state number when energy is in $E\sim E+\td E$. Furthermore,
\begin{equation}\label{intforfTFD}
  Z(\beta+it)=N_0\int_0^\infty e^{S(E)-\beta E}e^{-itE}\td E \,,
\end{equation}
where we used the relation between $N(E)$ and the entropy $S(E)$, $N(E)=N_0e^{S(E)}$ with a finite constant $N_0$.

In CFTs or AdS black holes {$S(E)-\beta E<0$ as $E\rightarrow\infty$}, so the integration of $|e^{S(E)-\beta E}|$ on $E\in(0,\infty)$ is finite. Then the Riemann-Lebesgue lemma says that
\begin{equation}\label{CFTZbetat1}
  \lim_{t\rightarrow\infty}Z(\beta+it)=\lim_{t\rightarrow\infty}N_0\int_0^\infty e^{S(E)-\beta E}e^{-itE}\td E=0\,,
\end{equation}
so
\begin{equation}
\lim_{t\rightarrow\infty} F(t)=0\,.
\end{equation}
The zero inner product at late time limit means that the complexity between $|\text{TFD}(t)\rangle$ and $|\text{TFD}(0)\rangle$ will grow forever. It will be model-dependent about how $F(t)$ decay to zero. It is compatible to the predictions in holographic conjectures. For some special models, if $F(t) \sim e^{-t}$ it will yield the linear-T complexity.

\section{Conclusions}\label{summ}
For the quantum circuits in quantum computation science, which we call `real quantum circuits', the complexity or operators or between two quantum states in general is not invariant under the unitary transformations of the states. We  call it `non-unitary invariant complexity'.
Towards understanding the complexity in quantum field theory rather than in real quantum circuits, many research have beed done based on intuitions from real quantum circuits. In particular, the non-unitary complexity.
Even though many interesting results have been reported in those research, we find that it may be a crucial question asking whether the complexity in quantum mechanics/field theory should be non-unitary invariant or not.

To answer this question, we tried to check if the `non-unitary invariant' property of the complexity can be compatible with the general framework of quantum mechanics/field theory.
We find, if the complexity is non-unitary invariant, there may be some problems: four issues are summarized in section \ref{shortsummary1}.  Here, we repeat them for readers' convenience.
\begin{enumerate}
\item[(1)] For a given physical situation, there are two ways to define the complexities $\C_r$ and $\C_l$. If the complexity is non-unitary-invariant, in general, $\C_r$ and $\C_l$ may be different and can not tell us the same physics, but there is no good physical reason to tell which one is correct.
\item[(2)] If the complexity is non-unitary-invariant, there are too many free parameters in the theory. Along this line, current studies are based on some artificial choices of the parameters, which make intrinsic understanding of the complexity difficult.
\item[(3)] The non-unitary-invariant complexity may be in general in conflict with the fundamental method and symmetry of quantum physics based on the Lagrangian/Hamiltonian formalism because they suggest the complexity is unitary-invariant.
\item[(4)] The framework of the quantum field theory assumes that physical observables are encoded in the generating functional. However, the non-unitary-invariant complexity implies that the same generating functional can give different physics.
\end{enumerate}

We want to emphasize that all these problems do not arise in the complexity of real quantum circuits. They arise if we simply adopt the ``non-unitary-invariance'' of the complexity for quantum field theory.  Quantum mechanics/field theory are not a naive ``continuous version'' of real quantum circuits: (1) the quantum circuits and quantum mechanics/field theory have many essential differences; (2) some properties are true in quantum circuits but may not be true in quantum mechanics/field theory, and vice versa.
We find that the `non-unitary-invariance' of the complexity may not be  compatible with quantum mechanics/field theory, contrary to the real quantum circuits. We also argue that  the locality should be defined in an intrinsic way.  Such an intrinsic locality is unitary invariant and is consistent with the unitary invariant complexity.

To resolve the above problems, {we propse that the complexity of operators and between quantum states should be unitary invariant. For quantum states, we proposed a deformed complexity} formula between states in quantum mechanics/field theory
\begin{equation}\label{defholoCeq2b}
  \bar{\C}(|\psi_1\rangle,|\psi_2\rangle)=-\ln|\langle\psi_1|\psi_2\rangle|^2\,.
\end{equation}
which comes from the extensive property of the complexity based on the volume dependence of the holographic complexity.
This formula looks very simple but its contents are indeed rich.

First, we have shown that the complexity between the ground state of a given Hamiltonian and the eigenstate of the field operator is given by the partition function. In the classical limit, the partition function is given by the on-shell Euclidian action, by which we gave a proof for the relationship between the ``path-integral complexity'' and the Liouville action.

Second, we also used our proposal to give natural explanations for both the CV and CA conjectures and clarified the reference and target states of them. The CV conjecture is dual to the complexity between the TFD state and its perturbed TFD state under a marginal operator. The CA conjecture is dual to the complexity between a TFD state and the eigenstate of the field operator of the boundary field theory. This difference between the CV and CA conjecture from field theory perspective naturally explains why the holographic CV and CA conjecture show different time evolution.

Third, we apply our proposal to chaotic systems.  We have  shown that following suggested behaviors of the time-dependent complexity can be reproduced by our proposal:

\begin{enumerate}
\item[(1)] The complexity grow linearly at late time if the classical system is chaotic. The growth rate is determined by the {\it Lyapunov exponent}
\begin{equation}\label{complexL1100}
  \bar{\C}(t)={\lambda}_Lt+\cdots\,.
\end{equation}
\item[(2)]
This linear-in-time behavior breaks down at the time scale $t \sim t_{cl}$
\begin{equation}\label{logtime1100}
t_{cl}:=-\frac{1}{2{\lambda}_L}\ln(c_1\hbar) \,,
\end{equation}
where $c_1$ is model-dependent constant.
\item[(3)]  For field theory systems Eq.\eqref{complexL1100} can be understood as
\begin{equation}\label{complexL21100}
  \dot{\bar{\C}}(t)\propto E+\cdots\,,
\end{equation}
which is similar to the growth rate of the complexity in the CV and CA conjectures.
\end{enumerate}

Finally, we have made two comments on our proposal. i) Our proposal can  resolve the problems in the Fubini-Study distance and ii) if it is applied to the TFD states it is compatible with the holographic complexities at late time.

It is often claimed~\cite{Brown:2017jil, Balasubramanian:2019wgd} that the complexity must be non-unitary invariant because a unitary-invariant complexity cannot reproduce the ``expected'' time evolution of the complexity: for a chaotic system with $N$ degrees of freedom, the complexity evolves as time goes in three stages: linear growth until $t\sim e^{N}$, saturation and small fluctuations after then, and quantum recurrence at $t\sim e^{e^N}$.
However, we provide counter examples of this claim \cite{Yang:2019iav}, where the unitary-invariant or bi-invariant complexity can indeed realize the expected time evolution. The examples in \cite{Yang:2019iav} support our claim that the complexity may be unitary-invariant!

We want to emphasize again that there is nothing wrong with the ``non-unitary-invariance'' of the complexity in {\it real quantum circuits}. The essential point we want to make is that there is no good reason to assume ``non-unitary-invariance'' for {\it quantum mechanics/field theory.} Rather, we find that there are some conflicts with the framework of quantum mechanics/field theory.  By presenting several interesting and consistent results based on a unitary-invariant complexity formula, we want to demonstrate that the {\it unitary-invariant} complexity can be still valid in the case of quantum mechanics/field theory.

\acknowledgments
The work of K.-Y. Kim was supported by Basic Science Research Program through the National Research Foundation of Korea(NRF) funded by the Ministry of Science, ICT $\&$ Future Planning(NRF- 2017R1A2B4004810) and GIST Research Institute(GRI) grant funded by the GIST in 2019. C. Niu is supported by the Natural Science Foundation of China under Grant No. 11805083. C.Y. Zhang is supported by Project funded by China Postdoctoral Science Foundation. We also would like to thank the APCTP(Asia-Pacific Center for Theoretical Physics) focus program,``Holography and geometry of quantum entanglement'' in Seoul, Korea for the hospitality during our visit, where part of this work was done.

\bibliographystyle{JHEP}
\bibliography{FG-ref-state}

\end{document}